\documentclass[longauth]{aa}
\usepackage{float}
\usepackage{placeins}
\usepackage{graphicx,url,twoopt,natbib}
\usepackage{xspace}
\usepackage{multirow}
\usepackage{hyperref}
\usepackage[usenames,dvipsnames,svgnames,table]{xcolor}
\hypersetup{
    colorlinks=true,
    linkcolor={blue},
    citecolor={blue},
    urlcolor={blue}
}
\usepackage[flushleft]{threeparttable}

\newcommand{\msun}{{$M_\odot$}}

\newcommand{\OII}{[\textrm{O}~\textsc{ii}]\xspace}
\newcommand{\OIII}{[\textrm{O}~\textsc{iii}]\xspace}
\newcommand{\OI}{[\textrm{O}~\textsc{i}]\xspace}

\newcommand{\Ha}{\textrm{H}\ensuremath{\alpha}\xspace}
\newcommand{\Hb}{\textrm{H}\ensuremath{\beta}}
\newcommand{\Hg}{\textrm{H}\ensuremath{\gamma}\xspace}

\usepackage{graphicx}
\usepackage{txfonts}
\defcitealias{Curti_20}{\scshape C20}
\defcitealias{Chakraborty_24}{\scshape C25}
\defcitealias{Bian_18}{\scshape B18}
\defcitealias{Sanders_24}{\scshape S24}
\defcitealias{Nakajima_22}{\scshape N22}
\defcitealias{Izotov_19}{\scshape I19}
\usepackage{orcidlink}
\usepackage{lastpage}
\usepackage{float}

\begin{document} 

   \title{Insights on Metal Enrichment and Environmental Effect at $z\approx5-7$ with JWST ASPIRE/EIGER and Chemical Evolution Model}

\author{Zihao Li\inst{1,2,3}\fnmsep\thanks{\email{zihao.li@nbi.ku.dk}}\orcidlink{0000-0001-5951-459X},
        Koki Kakiichi\inst{1,2}\orcidlink{0000-0001-6874-1321},
        Lise Christensen\inst{1,2}\orcidlink{0000-0001-8415-7547},
        Zheng Cai\inst{3}\orcidlink{0000-0001-8467-6478},
        Avishai Dekel\inst{4,5}\orcidlink{0000-0003-4174-0374},
        Xiaohui Fan\inst{6}\orcidlink{0000-0002-6336-3007},
        Emanuele Paolo Farina\inst{7}\orcidlink{0000-0002-6822-2254},
        Hyunsung D. Jun\inst{8}\orcidlink{0000-0003-1470-5901},
        Zhaozhou Li\inst{4}\orcidlink{0000-0001-7890-4964},
        Mingyu Li\inst{3}\orcidlink{0000-0001-6251-649X},
        Maria Pudoka\inst{6}\orcidlink{0000-0003-4924-5941},
        Fengwu Sun\inst{9}\orcidlink{0000-0002-4622-6617},
        Maxime Trebitsch\inst{10}\orcidlink{0000-0002-6849-5375},
        Fabian Walter\inst{11}\orcidlink{0000-0003-4793-7880},
        Feige Wang\inst{12}\orcidlink{0000-0002-7633-431X},
        Jinyi Yang\inst{12}\orcidlink{0000-0001-5287-4242},
        Huanian Zhang\inst{13}\orcidlink{0000-0002-0123-9246},
        Siwei Zou\inst{14,15}\orcidlink{0000-0002-3983-6484}
}

   \institute{Cosmic Dawn Center (DAWN), Denmark
         \and
             Niels Bohr Institute, University of Copenhagen, Jagtvej 128, DK2200 Copenhagen N, Denmark
        \and Department of Astronomy, Tsinghua University, Beijing 100084, China
        \and Center for Astrophysics and Planetary Science, Racah Institute of Physics, The Hebrew University, Jerusalem, 91904, Israel
        \and Santa Cruz Institute for Particle Physics, University of California, Santa Cruz, CA 95064, USA
        \and Steward Observatory, University of Arizona, 933 North Cherry Avenue, Tucson, AZ 85721-0065, USA
        \and International Gemini Observatory/NSF NOIRLab, 670 N A’ohoku Place, Hilo, Hawai'i 96720, USA
        \and Department of Physics, Northwestern College, 101 7th Street SW, Orange City, Iowa 51041, USA
        \and Center for Astrophysics $|$ Harvard \& Smithsonian, 60 Garden St., Cambridge, MA 02138, USA
        \and LUX, Observatoire de Paris, Université PSL, Sorbonne Université, CNRS, 77 Av. Denfert-Rochereau, F-75014 Paris, France
        \and Max Planck Institut f\"ur Astronomie, K\"onigstuhl 17, D-69117, Heidelberg, Germany
        \and Department of Astronomy, University of Michigan, 1085 S. University Avenue, Ann Arbor, MI 48109, USA
        \and Department of Astronomy, Huazhong University of Science and Technology, Wuhan, Hubei 430074, China
        \and Chinese Academy of Sciences South America Center for Astronomy, National Astronomical Observatories, CAS, Beijing 100101, China
        \and Departamento de Astronom\'ia, Universidad de Chile, Casilla 36-D, Santiago, Chile
            }

   % \date{Received September 15, 1996; accepted March 16, 1997}

% \abstract{}{}{}{}{} 
% 5 {} token are mandatory

  \abstract
    {We present the mass-metallicity relation (MZR) for a parent sample of 604 galaxies at $z=5.34-6.94$ with \OIII\ doublets detected, using the deep JWST/NIRCam wide field slitless spectroscopic (WFSS) observations in 26 quasar fields. The sample incorporates the full observations of 25 quasar fields from the JWST Cycle 1 GO program ASPIRE and the quasar SDSS J0100+2802 from the JWST EIGER program.
    We identify 204 galaxies residing in overdense structures using the friends-of-friends (FoF) algorithm. We estimate the electron temperature of $2.0^{+0.3}_{-0.4}\times10^4$ K from the \Hg\ and $\OIII_{4363}$ lines in the stacked spectrum, indicating a metal-poor sample with median gas phase metallicity 12+$\log(\mathrm{O/H})=7.65^{+0.26}_{-0.15}$.  
    With the most up-to-date strong line calibration based on NIRSpec observations, we find that the MZR shows a metal enhancement of $\sim0.2$ dex at the high mass end in overdense environments. However, compared to the local Fundamental Metallicity Relation (FMR), our galaxy sample at $z>5$ shows a metal deficiency of $\sim0.2$ dex relative to FMR predictions. We explain the observed trend of FMR with a simple analytical model, favoring dilution from intense gas accretion over outflow to explain the metallicity properties at $z > 5$. 
    Those high-redshift galaxies are likely in a rapid gas accretion phase, during which their metal and gas contents are in a non-equilibrium state. According to model predictions, the protocluster members are closer to the gas equilibrium state than field galaxies and thus have higher metallicity and are closer to the local FMR.
    Our results suggest that the accelerated star formation during protocluster assembly likely plays a key role in shaping the observed MZR and FMR, indicating a potentially earlier onset of metal enrichment in overdense environments at $z\approx5-7$. 
    }

               \titlerunning{Metal Enrichment and Environmental Effect at $z\approx5-7$}
               \authorrunning{Z. Li et al.}

   \maketitle
%
%-------------------------------------------------------------------

\section{Introduction}
The gas-phase metallicity (hereafter referred to as metallicity for simplicity) of galaxies represents the current state of chemical enrichment, bearing the imprints of both internal and external effects such as star formation, gas accretion, feedback, and mergers.
While the star-formation density peaks at cosmic noon \citep{2014ARA&A..52..415M}, 
studying galaxies at the epoch of reionization (EoR) provides us with a more complete picture of the earlier evolution of galaxies. 

The galaxies' metallicity is found to be highly correlated with their 
stellar masses (i.e., mass-metallicity relation, MZR) across a wide mass range 
($10^6-10^{11}$\msun) and a wide redshift range (from $z=0$ to $z\gtrsim6$) \citep{Tremonti_04,Maiolino_08,Andrews_13,Zahid_14,Curti_20,Sanders_21,Henry_21,Nakajima_23,Langeroodi_23,Stephenson_24,Pallottini_24}. The origin of MZR is still under debate. Many works have explored the various origins of the MZR, such as inflow and outflow properties \citep{Erb_06,Finlator_08,Spitoni_10,Bassini_24,Perez-Diaz_24}, secular evolution \citep{Somerville_15}, stellar age \citep{Duarte_22}, and AGN activity \citep{Thomas_19}.
A common interpretation within the outflow and inflow scenarios is that outflow efficiency decreases with increasing gravitational potential, thus increasing the metal retention for massive galaxies \citep{Finlator_08}.
In contrast, \citet{Baker_23a} argued that the MZR arises because the stellar mass is proportional to the total metal production in a galaxy, rather than due to more massive galaxies retaining metals more efficiently.

On the other hand, observations reveal that the MZR is not universal and can vary with environments. This influence on galaxy evolution is known as environmental effects, in many different forms such as ram pressure stripping and strangulation \citep{McCarthy_08, Peng_15, Boselli_22,Xu_25}, merger \citep{Brennan_15,Delahaye_17}, harassment \citep{Boselli_06,Bialas_15}, and preprocessing \citep{Fujita_04,Werner_22,Lopes_24}. The local galaxy clusters host galaxy populations with varied properties \citep{Murphy_09,Bahe_13}, and such environmental effects are also found at higher redshifts up to $z\gtrsim2$ \citep{Hatch_17,Tadaki_19,Namiki_19,Wang_22,Lemaux_22,Liu_23,Perez-Martinez_23,Hughes_24,Forrest_24}.

However, the environmental impact on the MZR at high redshifts—particularly within the first gigayear after the Big Bang—remains uncertain, with mixed observational evidence and different theoretical interpretations.
While some studies find metal-deficient galaxies in overdense environments at $z\sim2$, such as \citet{Valentino_15} in the CL J1449+0856 cluster at $z=1.99$, \citet{Li_22} and \citet{Wang_22} in the BOSS1244 protocluster at $z=2.24$, and \citet{Perez-Martinez_24} at $z=2.53$, others report metal-enhancement in protoclusters in similar redshifts. For example, \citet{Shimakawa_15} found the MZR to be systematically shifted upward by $\gtrsim0.1$ dex in two rich protoclusters compared to coeval field galaxies, and \citet{Shimakawa_18} reported enhanced galaxy formation in the densest regions of a protocluster at $z=2.5$. \citet{Perez-Martinez_23} found mild metal enhancement in the Spiderweb protocluster at $z=2.16$, consistent with a scenario of efficient gas recycling and confinement due to the external intracluster medium (ICM) pressure. Similarly, \citet{Adachi_25} found $0.08-0.15$ dex metal enhancement in the X-ray cluster XCS2215 at $z=1.46$, and they also attribute this to the confinement of gas by ICM pressure. 

Conversely, \citet{Calabro_22} observed a metallicity deficit of $\sim$0.1 dex in overdense regions at $2<z<4$, and \citet{Sattari_21} also found metal-deficient populations in dense environments. 
Some studies also suggest less environmental impact on the MZR. For instance, \citet{Namiki_19} found no significant differences in the MZR for a cluster at $z=1.52$ compared with the field. However, the observed differences reported in those works may well arise from measurement uncertainties, as the typical offsets of $\sim 0.1$ dex are comparable to the uncertainties.

Hydrodynamical simulations have also started to explore this complexity \citep{Fukushima_23,Esposito_25,Morokuma_25}. For example, \citet{Wangk_23} employed the EAGLE simulation \citep{Crain_15} and reported strong environmental influences on the MZR at both $z = 0$ and $z > 2$, attributing to both gas accretion and gas stripping. These results suggest that metallicity evolution may depend on both local physical processes, such as cold-mode accretion-driven gas dilution \citep[e.g.,][]{Wang_22}, and external processes, such as recycling-driven enrichment due to external pressure \citep[e.g.,][]{Perez-Martinez_23}. Additional high-redshift observations and simulations are needed to fully understand these environmental influences.

With the recent surge in observations from JWST, an increasing number of high-redshift galaxies have been identified during the Epoch of Reionization (EoR). This advancement enables the study of early galaxies using large statistical samples in both field and protocluster environments.  The JWST ASPIRE program (A SPectroscopic survey of biased halos In the Reionization Era; program ID 2078, PI: F. Wang) aims to target 25 quasars at $z>6.5$ using JWST NIRCam/WFSS with filter F356W and meanwhile search for \OIII+ \Hb\ emitters at $z=5.34-6.94$ \citep{Wangf_23}. The JWST EIGER program (Emission-line galaxies and the Intergalactic Gas in the Epoch of Reionization; program ID 1243, PI S. Lilly) targets 6 quasar fields, each centered on a quasar at $5.98<z<7.08$ \citep{Kashino_23}. Both ASPIRE and EIGER utilize the F356W filter for NIRCam/WFSS observations, providing identical redshift coverage.
Those observations allow us to identify galaxies in both overdense and field environments in a large survey volume, and reveal the impacts of environments on galaxy formation and evolution at the EoR. Recent JWST observations have unveiled an abundant population of high-redshift overdensities at $z\sim5-6$ through the detection of either \Ha or \OIII emissions with NIRCam/WFSS or NIRSpec \citep{Laporte_22,Wangf_23,Castellano_23,Kashino_23,Helton_24,Helton_24b,Sun_24,Herard-Demanche_25,Champagne_25}. \citet{Helton_24} discovered older stellar populations in a $z=5.4$ protocluster with JWST, pointing to earlier star formation and mass assembly in protoclusters, and \citet{Champagne_25} also observed galaxies in a protocluster at $z=6.6$ being more massive and older than field galaxies. Similarly, \citet{Morishita_25c} found more evolved stellar populations at $z\sim5.7$, in agreement with accelerated star formation.  Furthermore, \citet{Morishita_23} identified a galaxy protocluster at the redshift $z = 7.9$, using JWST NIRSpec, and a significant metallicity scatter was observed within this system, suggesting a rapid gas cycle in overdense regions \citep{Morishita_25}. \citet{Liq_25} observed enhanced dust extinction, weaker Lyman-alpha emission, and/or higher damped Lyman-alpha absorption in protocluster galaxies at redshifts $4.5 < z < 10$. 

In this paper, we present the first effort to investigate MZR at $5<z<7$ in both blank fields and overdense environments, utilizing a large, homogeneous sample selected through JWST NIRCam/WFSS observations in the F356W filter.

This paper is organized as follows. In Section \ref{sec:data}, we describe the observations and data used in our analysis. In Section \ref{sec:method}, we present the methodology, including emitter finding and spectra stacking. In Section \ref{sec:MZR} and \ref{sec:FMR}, we present our main results on MZR and FMR. We introduce a simple analytical model in \ref{sec:model}, and we compare our observations with analytical model and simulations in \ref{sec:discussion}. In Section \ref{sec:summary}, we summarize our results and their implications.

Throughout this article, we adopt the AB magnitude system, and assume a flat $\Lambda$CDM cosmology with $\Omega_m=0.3,\ \Omega_\Lambda=0.7$, and $H_0=70\ {\rm km\ s^{-1} Mpc}^{-1}$.

%--------------------------------------------------------------------
\section{Observations and data reduction} \label{sec:data}
\subsection{NIRCam imaging data reduction}
The images are processed in the same way as described in \citet{Wangf_23}.
Briefly, the reduction was performed using version 1.8.3 of the JWST Calibration Pipeline with the reference files from version 11.16.15 of the standard Calibration Reference Data System. The 1/f noise is removed on a row-by-row and column-by-column basis. We then created a master median background for each combination of detector and filter using all calibrated exposures from stage 2. These master backgrounds were subsequently scaled and subtracted from the individual exposures to remove extra detector-level noise.
After Stages 2 and 3, the images are aligned to Gaia DR3 and drizzled to a common pixel scale of 0\farcs031/pixel.
\subsection{NIRCam WFSS data reduction}
We use version 1.13.4 of the JWST Calibration pipeline \texttt{CALWEBB} Stage 1 to calibrate individual NIRCam WFSS exposures, with reference files \texttt{jwst\_1321.pmap}. The $1/f$ noise is then subtracted along rows for Grism-R exposures using the routine described in \citet{Wangf_23}. The world coordinate system (WCS) information is assigned to each exposure with the \texttt{assign\_wcs} step. The flat field is done with \texttt{CALWEBB} stage-2. We build the median backgrounds based on all of the WFSS exposures, which are then scaled and subtracted from each individual exposure. The astrometric offsets are measured between each of the short wavelength (SW) images and the fully calibrated F356W mosaic to align each grism exposure with the direct image. The WCS alignment ensures the tracing model works properly.

The pre-processed WFSS exposures are then processed by the Grism Redshift \& Line {Analysis tool} (\textsc{Grizli}; \citealt{Brammer_2022}). We use the V9 spectral tracing and grism dispersion models\footnote{\url{https://github.com/npirzkal/GRISM_NIRCAM}}.
The detection catalogs for spectral extraction are built from the F356W direct image, and the continuum cross-contamination is subtracted by \textsc{Grizli} forward modeling using the F356W image as the reference image for each grism exposure. The 1D spectra are extracted with optimal extraction \citep{Horne_86} with optimal profiles generated from F356W images and corresponding segmentation maps.

\section{Methodology and measurements} \label{sec:method}
\subsection{Emitter and overdensity identification}

\begin{figure}[!t]
    \centering
    \includegraphics[width=0.48\textwidth]{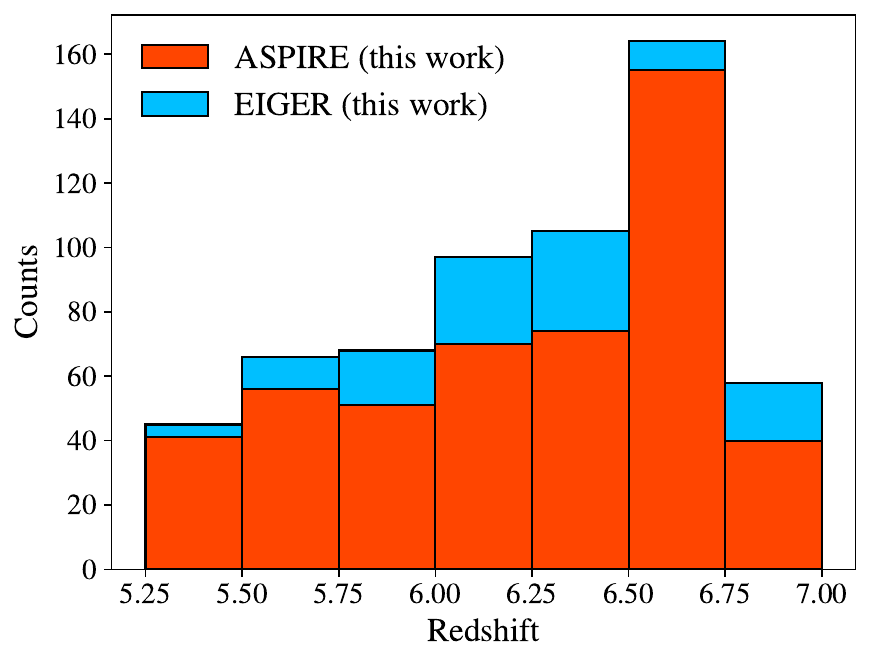}
    \caption{Redshift distribution of the full sample in this work. \OIII emitters from ASPIRE and EIGER programs are marked by red and blue, respectively. The EIGER sample is stacked on top of the ASPIRE sample.}
    \label{fig:z_dis}
\end{figure}

To search for \OIII\ emitters, we use the automatic line searching algorithm detailed in \citet{Wangf_23} on continuum-subtracted spectra. In brief, we applied a peak finding algorithm to search for all pixels with $\rm S/N>3$ and rejected peaks with two neighboring pixels with $\rm S/N<1.5$ to avoid likely hot pixels or cosmic ray signals. We then performed a Gaussian fitting to the remaining peaks, and accepted lines with $\rm S/N>5$ and FWHM wider than half the spectral resolution and narrower than seven times the spectral resolution. For each detected lines, we assumed that it is $\rm \OIII\lambda5007\text{\AA}$, and measure the $\rm S/N$ at the expected wavelength of $\rm \OIII\lambda4959\text{\AA}$ 
with the same FWHM as $\rm \OIII\lambda5007\text{\AA}$. We regard the \OIII\ emitter candidates as good if the significance of detection is over $2\sigma$. We then apply a color excess cut from broad band photometry ($\rm mag_{F200W}-mag_{F356W}>0.2$) to further reject candidates that are not likely to be \OIII\ emitters. We then visually inspect the remaining candidates. We compared the source morphology on direct image and emission lines on 2D spectra, and rejected sources whose emission lines' shapes are apparently different from the direct image, which are likely to be contaminated by other sources. We also visually check all the sources along the dispersion direction of the candidates to avoid contamination from nearby sources. We finally selected 487 \OIII\ emitters from 25 fields in ASPIRE. The emitter selection criteria for the EIGER field are outlined in \citet{Kashino_23}, and we use the publicly available catalog of 117 \OIII emitters provided by the EIGER collaboration\footnote{\url{https://eiger-jwst.github.io/data.html}}. The spectra of \OIII emitters are extracted from our reduced products in the SDSS J0100+2802 field, using coordinates from the EIGER collaboration. Finally, the complete sample incorporates 604 \OIII emitters in total. The redshift distribution is shown in Fig. \ref{fig:z_dis}.

Following \citet{Helton_24}, we use a Friends-of-Friends (FoF) algorithm to identify overdense structures. This algorithm selects galaxy groups by searching for companions around a central galaxy within a projected separation $d_{\text{link}}=500\ \rm{kpc}$ and line-of-sight (LOS) velocity $\sigma_{\text{link}}=500\ \rm{km/s}$. The algorithm iteratively performs the process above for all the companions identified until no more galaxies are found. Similar to \citet{Helton_24}, we require a minimum number of constituent galaxies $(N_{\rm galaxies} \ge 4)$ as a cluster. We note that these overdensities are not ``real'' clusters under the standard definition, which describes mature, gravitationally bound structures in virial equilibrium. They may not necessarily be protoclusters either, as it is uncertain whether these systems will ultimately evolve into clusters at $z=0$. We refer to them as "clusters" throughout this work for simplicity, as they exhibit greater clustering compared to isolated galaxies. In addition, as protoclusters can span scales of $\sim$deg \citep{Hung_25}, our selected overdensities on scales of $\sim$arcmin in each field may only represent some sub-components of a protocluster, without knowing the exact location of the main halo.

We finally identified 204 galaxies in overdense environments (cluster) and 400 galaxies that are not linked to another (field). Among the selected clusters, an extreme overdensity at $z=6.6$ in J0305--3150 with $\delta_\text{gal}=12.6$ has been reported in \citet{Wangf_23}. And three major overdensities at $z\approx6.2,6.3,6.8$ in the SDSS J0100+2802 field have also been reported in \citet{Kashino_23}.

\subsection{SED fitting}
The spectral energy distribution (SED) modeling of our sample is carried out using the Bayesian code \texttt{BEAGLE} \citep{Chevallard_16}, incorporating broadband photometry and \OIII\ line fluxes as inputs. 
\texttt{BEAGLE} computes the stellar and nebular emissions based on an updated version of the \citet{Bruzual_03} stellar population synthesis models \citep{Gutkin_16}.
We adopt a delayed-$\tau$ star formation history (SFH), the Small Magellanic Cloud (SMC) dust attenuation law, and a Chabrier initial mass function (IMF) \citep{Chabrier_03} with an upper mass limit of 100\msun. 
We note that the fitted parameters are sensitive to underlying assumptions, such as SFH, and we discuss the impact of these systematic uncertainties in Section~\ref{sec:caveats}.

Flat priors in log-space are applied for the characteristic star formation timescale, $\tau$, ranging from $10^7$ to $10^{10.5}$ years, and for stellar masses, spanning $10^4M_\odot$ to $10^{12}M_\odot$. 
We place log-uniform priors on stellar metallicity $\log(Z/Z_\odot)$ from -3 to 0, and ionization parameter $\log(\rm U)$ from -4 to -1. The metallicity and ionization parameter ranges are sufficiently broad to encompass the typical values suggested by relevant observations \citep{Strom_18,Reddy_23,Trump_23,Sanders_24}.
\texttt{BEAGLE} assumes that total interstellar metallicity, including dust and gas-phase metallicity, is the same as stellar metallicity.
The optical depth in the $V$ band is allowed to vary between 0 and 0.5 in log-space. For both the ASPIRE and EIGER fields, we utilize data from three JWST NIRCam bands: F115W, F200W, and F356W.

\subsection{Spectra stacking}
\begin{figure*}[t!]
    \centering
    \resizebox{18cm}{!}{\includegraphics{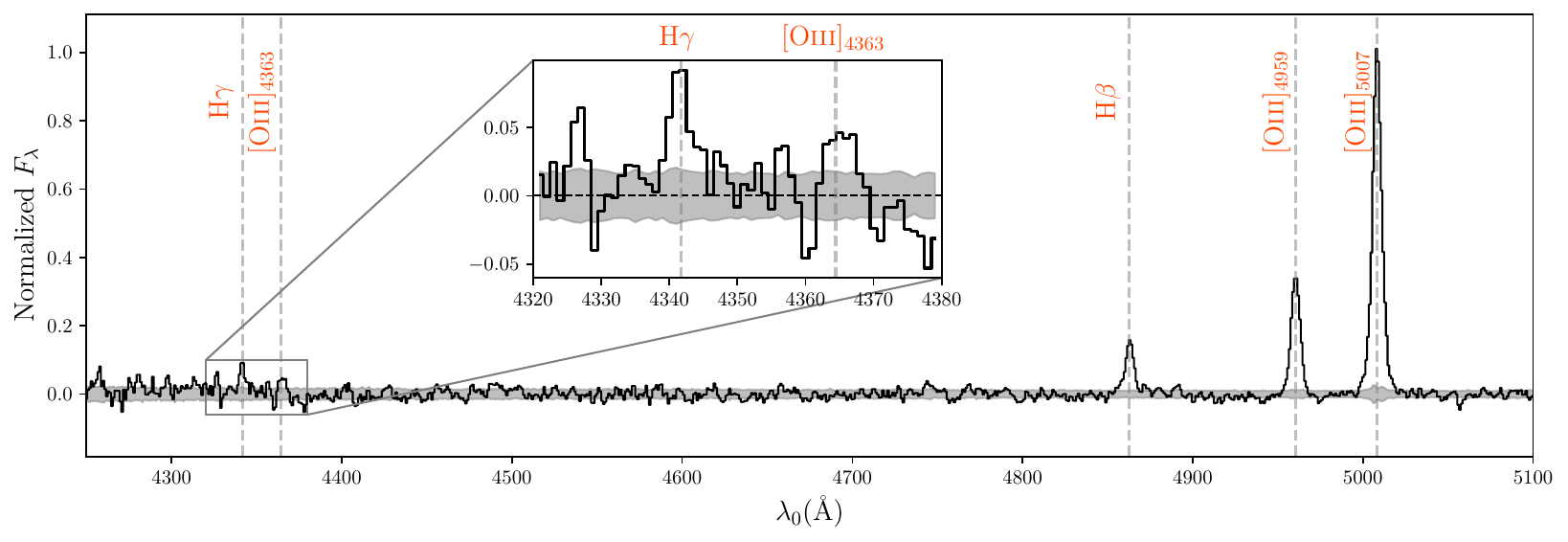}}
    \caption{Median stacked 1D rest-frame spectra for our full sample galaxies at $z>6.25$, with continuum subtracted. The fluxes are normalized by the peak of \OIII$_{5007}$ flux after stacking.}
    \label{fig:spec_stack}
\end{figure*}
To enhance the spectral signal-to-noise ratio (SNR), we stack galaxies that are expected to share similar metallicities and, consequently, similar line ratios. Given the strong correlation in the MZR \citep[e.g.,][]{Nakajima_23,Chakraborty_24,Sarkar_25}, it is reasonable to assume that galaxies with similar stellar masses have similar metallicities.
We divided the galaxy sample into two categories: overdense regions (clusters) and blank fields (fields), further stratified into three mass bins. The mass bins were uniformly spaced to ensure comparable SNRs in the resulting stacks for each bin. Before stacking, we subtract a polynomial continuum from each spectrum, removing potential background contamination. We resample our spectra to rest-frame on a common 1 \AA\ wavelength grid with flux preserved using \texttt{spectres} \citep{2017arXiv170505165C}. Following \citet{Wang_22}, to avoid the excessive weighting towards bright sources with stronger line fluxes, we normalized each spectrum by its measured \OIII\ flux. We take the median value of the normalized spectra at each wavelength grid, and the uncertainty is estimated by measuring the standard deviation from 1000 bootstrap realizations of the sample. 

The \Hg\ and $\OIII_{4363}$ emission lines at $6.25 < z < 6.95$ are redshifted into the F356W filter coverage. Fig. \ref{fig:spec_stack} presents the median-stacked 1D rest-frame spectrum for all galaxies within this redshift range. In the stacked spectrum, we detect the $\OIII_{4363}$ auroral line at a $2\sigma$ significance level, enabling us to determine the median metallicity of the full sample using the $T_e$ method. 
We do not further measure $T_e$ for subsamples divided into mass bins or field/cluster bins, as these subsamples span redshift ranges broader than $6.25 < z < 6.95$, resulting in incomplete spectral coverage. The reduced sample sizes in these bins hinder reliable $\OIII_{4363}$ measurements; consequently, any observed differences between the samples are more likely driven by random noise than by environmental effects.
The analysis is further discussed in Section \ref{sec:te}. 

\subsection{Flux measurements}

For individual targets, we apply \textsc{Grizli} to measure line fluxes by forward modeling. This process starts with the construction of a one-dimensional (1D) spectrum, which includes multiple Gaussian-shaped emission lines (\OIII doublets and \Hb) at a given redshift, combined with a continuum derived from sets of empirical spectra \citep{Brammer_08}. This modeled 1D spectrum is then used to generate a two-dimensional (2D) model spectrum, dispersing each pixel from the F356W direct image onto a grism detector frame for each exposure with grism sensitivity and dispersion function. 
The forward-modeled 2D spectrum is subsequently compared to the observational data, and a $\chi^2$-minimization is performed to identify the best-fit coefficients for both the continuum templates and the Gaussian amplitudes. A negative Gaussian amplitude is allowed to account for the effects of random noise. The best-fit emission-line fluxes are derived from the Gaussian components. Since we have required robust detection of \OIII lines in emitter identification, we have all $\text{SNR}_{\OIII}>3$ in the emitter identification algorithm. However, \Hb\ is not always detectable due to its faintness. In our subsequent measurements, we set an SNR cut of $\text{SNR}_{\Hb}>1.5$ to ensure spectra quality. In Appendix \ref{sec:bias}, we further quantify the bias of stacking low SNR spectra, leading to an overestimation of \OIII/\Hb\ ratio. 

After stacking, we measure line fluxes from the 1D spectra by fitting a combination of a polynomial continuum and multiple Gaussian emission line components. We note that the continuum is not necessarily physical, as the grism spectra are highly likely to be contaminated by light from nearby sources. Although \textsc{Grizli} has subtracted the contamination from bright sources, there are still possible residuals in the grism spectra due to potential tracing offsets. The polynomial continuum further subtracts residuals to ensure robust line flux measurements. We use two separate Gaussian components for the \OIII doublets and find that the line ratios of $\OIII_{5007}/\OIII_{4959}$ are close to $2.98:1$.
Line fluxes and uncertainties are derived from the uncertainties of their Gaussian components. 

\begin{figure}[t!]
    \centering
    \resizebox{8.5cm}{!}{\includegraphics{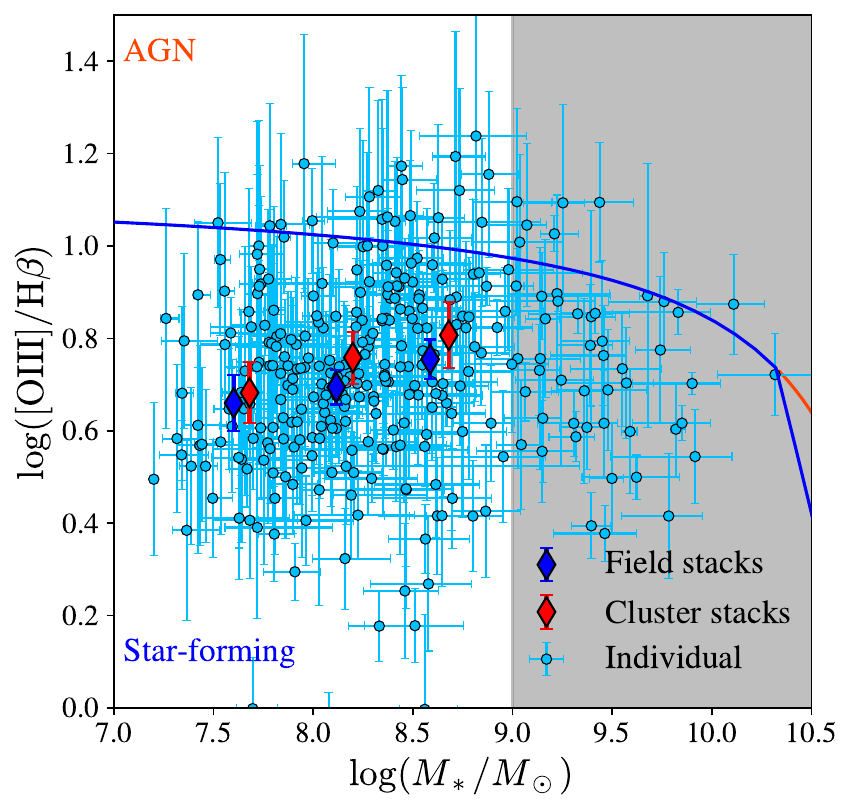}}
    \caption{Mass-excitation diagram for our sample galaxies. The blue circles represent individual measurements. The red and blue diamonds represent stacked measurements in cluster and field galaxies, respectively. The blue and orange curves indicate the lower and upper mass-excitation demarcation by \citet{Coil_15}. The AGNs (star-forming galaxies) are above (below) the demarcation. The gray shaded region indicates the high-mass end, which we did not analyze due to uncertainties in the metallicity calibrations in this regime.}
    \label{fig:MEx}
\end{figure}

In Fig.~\ref{fig:MEx}, we show the individual and stacked \OIII/\Hb ratios as a function of stellar mass, and compare them with the mass–excitation demarcation \citep{Coil_15}. We find that none of the galaxies in our sample exceed the AGN criteria at the $1\sigma$ confidence level. Thus, they are suitable for metallicity calibration for star-forming galaxies, as we will discuss in Section \ref{sec:MM}.

\subsection{Metallicity measurements}\label{sec:MM}

The direct electron temperature ($T_e$) method for determining metallicity relies on collisionally excited emission lines from metal ions. To estimate the electron temperature of doubly ionized oxygen ($\mathrm{O}^{++}$), the intensity ratio of $\OIII_{4363}$ to $\OIII_{4959,5007}$ is commonly used. The $\OIII_{4959}$ and $\OIII_{5007}$ lines originate from transitions from the $^1D$ energy level to the ground state $^3P$, while the $\OIII_{4363}$ line arises from a higher-energy transition from $^1S$ to $^1D$. The probability of excitation to the $^1D$ and $^1S$ states depends on the collisional rate for each ion, which is proportional to $N_e/v_e \propto N_e/T_e^{1/2}$, where $v_e$ represents the electron velocity. Therefore, measuring the emission intensities associated with these $\OIII_{4363}$ and $\OIII_{5007}$ transitions allows for an estimate of the electron temperature $T_e$. Likewise, the electron temperature of singly ionized oxygen ($\mathrm{O}^{+}$) can be estimated by the ratio of $\OII_{7322,7332}$ and $\OII_{3726,3729}$.

Given the electron temperature, metallicity is estimated using the direct method, based on physical conditions, emissivities, and observed fluxes, with photoionization models applied to account for the temperature and ionization structures of $\text{H}~\textsc{ii}$ regions \citep[e.g.,][]{Izotov_06,Perez-Montero_17,Amayo_21}.

Due to the challenges of detecting faint $\OIII_{4363}$, strong emission-line calibration methods have been widely used to determine the metallicity of high-redshift galaxies. These methods rely on the ratios between bright collisionally excited lines (e.g., $\OIII, \OII$) and Balmer recombination lines (e.g., \Ha, \Hb) to establish metallicity calibrations \citep{Pagel_79}. Since these lines are among the strongest metal lines in the optical spectrum, they are applicable to galaxies across a broad range of redshifts and luminosities.  Strong-line calibrations are constructed using samples where \(\OIII_{4363}\) is detected, enabling the establishment of empirical relationships between metallicity measured via the direct \( T_e \) method and strong emission-line ratios (e.g., \(\OIII_{5007}/\Hb\), \(\OIII/\OII\)) \citep{Pettini_24}. For these calibrations to be reliable, the galaxies used in their derivation should have physical properties similar to those of the target galaxies. Typically, strong-line calibrations are based on galaxies selected from the BPT diagram \citep[][]{Baldwin_81} that exhibit detectable \(\OIII_{4363}\) emission \citep[e.g.,][]{Bian_18}. 
Therefore, when applying these calibrations, it is essential to ensure that the galaxies share similar ionization mechanisms. In such cases, they often appear in similar regions of the BPT diagram, as a consequence of emissions arising from similar ionizing scenarios.

To improve metallicity measurements of high-redshift galaxies, many strong-line calibrations use high-ionization galaxies in the local universe as high-redshift analogs, and establish empirical relations that represent the ionization conditions for high-redshift galaxies (\citealt{Bian_18}, hereafter, \citetalias{Bian_18}, \citealt{Izotov_19}, hereafter, \citetalias{Izotov_19}, \citealt{Nakajima_22}, hereafter, \citetalias{Nakajima_22}, \citealt{Curti_20}, hereafter, \citetalias{Curti_20}). However, these local analogs may not fully represent the ionization conditions of high-redshift galaxies. \citet{Kewley_13} found a significant evolution of emission line ratios toward higher excitation at high-redshift, which is also confirmed by recent JWST observations \citep{Shapley_23}. Thus, more reliable metallicity diagnostics are needed to improve the accuracy of metallicity measurements.

For our galaxy sample, we adopt the two most recent metallicity diagnostics (\citealt{Sanders_24}, hereafter, \citetalias{Sanders_24}, and \citealt{Chakraborty_24}, hereafter, \citetalias{Chakraborty_24}). Unlike calibrations that rely on local analogs of high-redshift galaxies (e.g., \citetalias{Bian_18, Nakajima_22, Curti_20}), these diagnostics are derived directly from high-redshift galaxies observed with JWST/NIRSpec. Specifically, they are based on a sample of galaxies at redshifts $z = 2-9$ with detected $\OIII_{4363}$, allowing for metallicity estimates using the direct $T_e$ method. This approach provides a more precise characterization of high-redshift galaxy properties, reducing potential biases introduced by local analog calibrations.
However, a limitation of the $\OIII/\Hb$ (R3) diagnostic is that it produces a double-branched solution for a given line ratio. Since $\OII_{3727}$ falls outside the spectral coverage of the F356W grism, we adopt the low-metallicity solution with $12+\log(\rm O/H)\lesssim7.9$. This assumption is supported by the fact that most galaxies with similar stellar masses at comparable redshifts ($z\geq6-7$) have been confirmed as metal-poor using the $T_e$ method \citep{Sanders_24,Curti_23,Nakajima_23,Trump_23,Jones_23,Chakraborty_24}.

\section{The metallicities and mass-metallicity relation} \label{sec:MZR}

\subsection{Direct $T_e$ metallicity and empirical calibrations}\label{sec:te}

\begin{table}
\renewcommand{\arraystretch}{1.3}
\caption{Measurements of the median stack spectra for full sample and subsets with $\OIII_{4363}$ and \Hg\ coverage at $z>6.25$.}
\label{tab:stack_info}
\centering
\begin{tabular}{ccc}
\hline\hline
Property & Full sample & $z>6.25$ \\
\hline
$R3$ & $6.05\pm0.25$ & $6.30\pm0.34$ \\
$^{a}F_{\rm H\beta}$ & $16.52\pm0.67$ & $15.88\pm0.83$ \\
$^{a}F_{\rm H\gamma}$ & -- & $6.97\pm1.16$ \\
$^{a}F_{\rm \OIII_{4363}}$ & -- & $2.84\pm0.90$ \\
$T\rm (\mathrm{O}^{++})/10^4K$ & -- & $2.03^{+0.33}_{-0.42}$ \\
12+$\rm \log(\mathrm{O}^{++})$ & -- & $7.55^{+0.22}_{-0.13}$ \\
12+$\log(\text{O/H})_{T_e}$ & -- & $7.65^{+0.26}_{-0.15}$ \\
12+$\log(\rm {O/H})_{S24}$ & $7.57^{+0.23}_{-0.16}$ & $7.63^{+0.21}_{-0.18}$ \\
12+$\log(\rm {O/H})_{C24}$ & $7.66^{+0.24}_{-0.13}$ & $7.74^{+0.19}_{-0.18}$ \\
\hline
\end{tabular}
\tablefoot{$^{a}$ Line intensities are normalized with respect to $F_{\OIII_{5007}}=100$.}
\end{table}
 
In Fig. \ref{fig:spec_stack}, we can detect $\OIII_{4363}$ and \Hg\ for median stacked spectra of sample galaxies at $z>6.25$, and we can estimate the metallicity using the $T_e$ method. We measure the Balmer ratio $\Hg/\Hb= 0.44\pm0.04$, consistent with a recent study in the EIGER field \citep{Matthee_23}. Assuming the SMC dust curve, we estimate the dust attenuation as $A_V=0.38^{+0.54}_{-0.38}$, suggesting moderate dust in our sample, albeit with considerable uncertainties.

Following \citet{Trump_23}, we measure the \OIII\ ratio with dust attenuation corrected as follows:
\begin{equation}\label{eq:o3}
    \frac{\OIII_{4363}}{\OIII_{4959+5007}}=\frac{\OIII_{4363}}{\rm H_\gamma}\frac{\rm H_\beta}{\OIII_{4959+5007}} \times(2.1)^{-1},
\end{equation}
where we apply the intrinsic Balmer decrement ratio $\rm H\beta/H\gamma=2.1$ with the assumption of Case B recombination for $T = 10^4K$ \citep{Veilleux_87}. Since the $\mathrm{O}^{++}$ electron temperature $T_e(\mathrm{O}^{++})$ is largely insensitive to electron density $n_e$, we assume an electron density of $n_e=300~\rm cm^{-3}$.
We determine the $\mathrm{O}^{++}$ electron temperature, $T_e(\mathrm{O}^{++}) = (2.0^{+0.3}_{-0.4}) \times 10^4$ K, using \texttt{PyNeb} \citep{2015A&A...573A..42L} with default \citet{Fischer_04,Storey_00} atomic databases. The doubly-ionized oxygen abundance, $12+\log(\mathrm{O}^{++}/\mathrm{H^+})$, is measured to be $7.56^{+0.28}_{-0.13}$. Since our spectrum lacks \OII\ coverage, we have to assume a $\OIII/\OII$ ratio. For example, \citet{Matthee_23} adopted $\OIII/\OII=8\pm3$ based on empirical scaling \citep{Katz_23}. This assumption is reasonable and has been validated by recent JWST NIRSpec observations of galaxies at similar redshifts (e.g., \citealt{Sanders_24} reported a median ratio $\sim9$). To assess its reliability for our sample, we tested the value using a fixed-point iteration method. We start by assuming an initial guess $\OIII/\OII = 10$, then calculate the $T_e$-based metallicity (discussed in the next paragraph) and derive the corresponding $\OIII/\OII$ ratio using the \OIII/\OII$-$metallicity calibration (O32) from \citetalias{Chakraborty_24}. We then use this updated $\OIII/\OII$ value as input for the $T_e$ metallicity method. Iterating this process yields a converged value of $\OIII/\OII = 8.3$, which is close to the value assumed by \citet{Matthee_23}.
Nevertheless, we note that the relation between $\OIII/\OII$ and metallicity has been observed to have large characteristic scatters at $z=2-9$ \citep[e.g.,][]{Sanders_24}, and the relation between ionization parameter as traced by \OIII/\OII and metallicity still lacks a clear consensus \citep[e.g.,][]{Ji_22}. We caution potential systematic uncertainty introduced by this assumption. We further tested a range of \OIII/\OII ratios $\sim4-35$, as observed by \citet{Sanders_24}, and found that the resulting metallicity difference between the maximum and minimum ratios is $\sim0.1$ dex.

The $\mathrm{O}^{+}$ electron temperature is approximated as $ T_e(\mathrm{O}^{+})=1.5\times10^4\rm{K}$ using the $T_e(\mathrm{O}^{+})-T_e(\mathrm{O}^{++})$ relation from \citet{Izotov_06}. We assume zero abundance of $\mathrm{O^{+++}}$, as the ionization parameter implied by our adopted \OIII/\OII\ ratio lies within a range where photoionization models \citep[e.g.,][]{Perez-Diaz_22} predict little to no high excitation $[\textrm{O}~\textsc{iv}]$ emission. We then calculate the total gas phase metallicity 12+$\log(\mathrm{O/H})_{T_e}=12+\log(\mathrm{O}^{++}/\mathrm{H^+}+\mathrm{O}^{+}/\mathrm{H^+})=7.65^{+0.26}_{-0.15}$. 
This approach, using observed line ratios, assumes a homogeneous ISM structure and temperature in regions of $\mathrm{O}^{++}$ and $\mathrm{O}^{+}$. We caution against potential bias from temperature fluctuations within ionization regions \citep[e.g., see][]{Cameron_23}.

We compare the direct $T_e$ metallicity with the results from \OIII/\Hb\ ratio based on empirical strong line calibrations \citepalias{Sanders_24,Chakraborty_24}. We do not correct for dust for \OIII/\Hb\ ratios. Since their wavelengths are close to each other, dust attenuation has a negligible impact on the line ratios.
For our $z>6.25$ sub-sample, we find metallicities of 12+$\log(\mathrm{O/H})_{\text{S24}}=7.63^{+0.21}_{-0.18}$ and 12+$\log(\mathrm{O/H})_{\text{C24}}=7.74^{+0.19}_{-0.18}$,  both of which are consistent within $1\sigma$ with the measurement obtained from the direct method, although \citetalias{Chakraborty_24} predicts $\sim0.1$ dex higher. The measurements of the median stack spectra are summarized in Table \ref{tab:stack_info}. From \OIII/\Hb\ measurements, we find that the higher redshift galaxies at $z>6.25$ are slightly more metal-rich than the full sample, contrary to the redshift evolution of gas-phase metallicity \citep{Sarkar_25}. This is likely related to the overdense environments around $z>6$ quasars. We can see a significant galaxy number density excess in $z=6.25-6.75$ bins in Fig. \ref{fig:z_dis}, coinciding with the redshift range where the quasars are located. Such an environmental effect is further revealed in Section \ref{sec:mzr}.

\subsection{Mass-metallicity relation (MZR)}\label{sec:mzr}

\begin{figure*}[t!]
    \centering
    \resizebox{14cm}{!}{\includegraphics{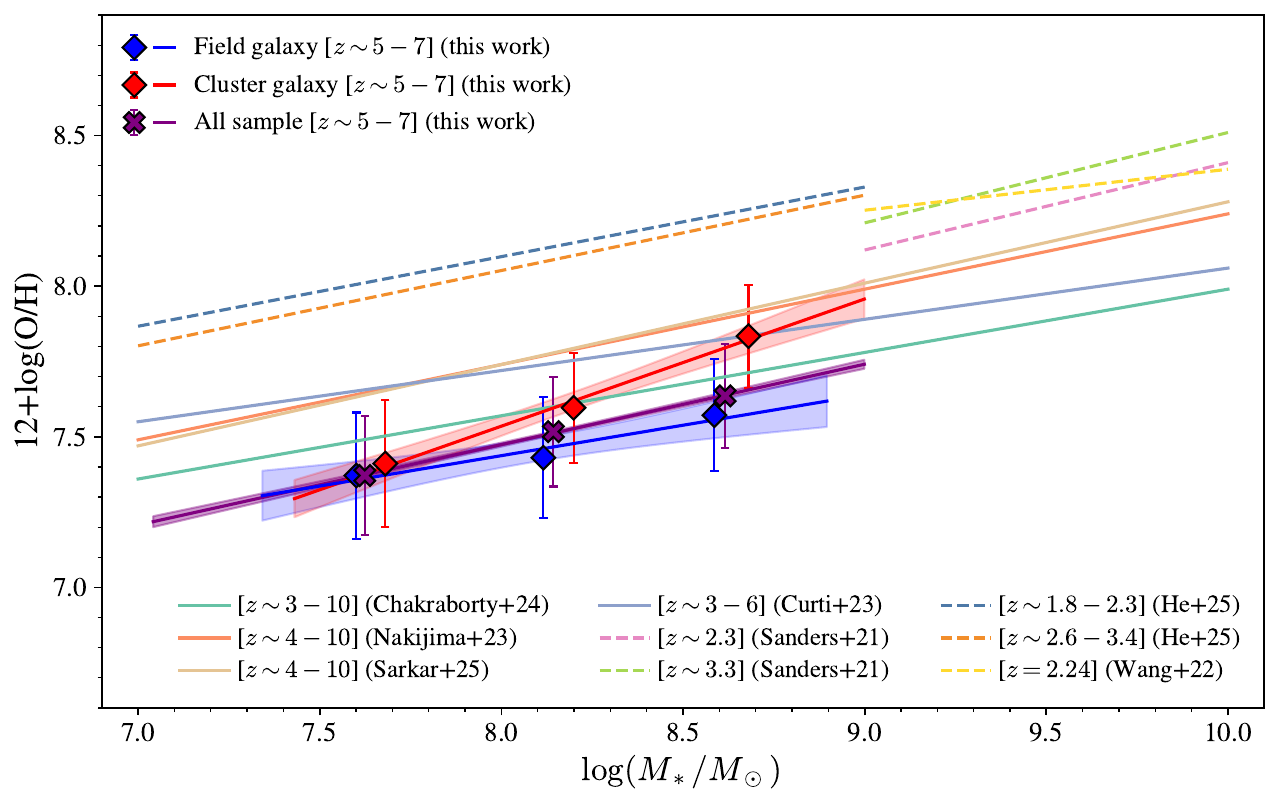}}
    \caption{The mass-metallicity relation for galaxies in protoclusters (red) and blank fields (blue), which are based on R3 calibration from \citetalias{Chakraborty_24}. The \citetalias{Sanders_24} calibration provides similar results, which are listed in the Table. \ref{tab:mzr_info} for comparison. 
    The results from other works in the literature are shown with different colors, which are also listed in the Table. \ref{tab:mzr_info} for reference. Higher redshift ($z>3$) measurements in literature are shown in solid lines, and lower redshift ($z\sim2-3$) measurements are shown in dashed lines.
    }
    \label{fig:mzr}
\end{figure*}

Since we use the R3 diagnostic to estimate MZR, we acknowledge the caveat of the R3 diagnostic that it provides two possible metallicity solutions for a given line ratio. In star-forming dominated systems, the \OIII/\OII ratio (O32), which is observed to decrease monotonically with increasing metallicity \citep[e.g.,][]{Maiolino_08,Nakajima_22,Sanders_24}, and is therefore often used to break degeneracies between different solutions \citep{Nakajima_23,Heintz_23,Sarkar_25}.

Without spectral coverage of \OII, it is challenging to distinguish between the two solutions. According to MZR at similar redshifts \citep{Chakraborty_24, Sarkar_25}, more massive galaxies $(M_*\gtrsim10^9M_\odot)$ are likely to lie on the high-metallicity branch of the calibration. The inclusion of high-mass galaxies could introduce uncertainties in metallicity estimation due to the confusion in line ratios.  We mitigate this issue by excluding the most massive galaxies ($\log(M_*/M_\odot)>9$) from MZR analysis and only applying the low-metallicity solution. The metallicities for these low-mass galaxies can be more safely estimated using the lower-branch solution.

We divide both our field and cluster samples into three mass bins. We stack these galaxy spectra with the method described in Section \ref{sec:method}.
We then determine the metallicity from the \OIII/\Hb\ ratio using empirical calibrations from \citetalias{Chakraborty_24}. 
% We present the mass-metallicity relation for low mass galaxies ($\log(M_*/M_\odot)<10^{9}$) in our sample.
To measure the MZR, we follow the parametrization used in \citet{Sanders_21}:
\begin{equation}
12 + \log(\text{O/H}) = \gamma \times \log \left( \frac{M_{*}}{10^{10}M_{\odot}} \right) + Z_{10},\label{eq:mzr}
\end{equation}
where the stellar mass is normalized with $10^{10}M_\odot$, and $\gamma$ and $Z_{10}$ are the slope and metallicity intercept. We present the MZR for both field and cluster galaxies in Fig. \ref{fig:mzr}, compared with literature studies. 

\subsubsection{MZR of the full sample}

We perform linear regression on the full sample using Eq. \ref{eq:mzr}. The high sensitivity of NIRCam grism allows us to measure low mass MZR down to $M_*\sim10^{7}M_\odot$. We determine the best-fit slope of $\gamma=0.26\pm0.01$ and an intercept of $Z_{10}=8.00\pm0.01$. Table \ref{tab:mzr_info} shows our measurements alongside those from the literature. 
In addition to \citetalias{Chakraborty_24} calibration, we also show the measurements using \citetalias{Sanders_24} calibration in Table \ref{tab:mzr_info} for comparison. We can see that the results based on \citetalias{Sanders_24} and \citetalias{Chakraborty_24} are consistent with each other.

We find that both the slope and intercept are reasonably consistent with \citet{Chakraborty_24}, who conducted the first MZR measurements using the direct $T_e$ method. Our best-fit MZR is $\sim0.2$ dex lower than those observed in CEERS \citep{Nakajima_23}, JADES \citep{Nakajima_23} and JADES+Primal \citep{Sarkar_25}. A similar offset has been found by \citet{Chakraborty_24}, who attributed their discrepancy to the systematic offset between the direct $T_e$ method and strong line calibration-based methods. Since we use \citetalias{Chakraborty_24} calibration derived from high-redshift direct $T_e$ method, our results are more consistent with \citet{Chakraborty_24}. \citet{Heintz_23} and also revealed a low metallicity intercept, who applied \citetalias{Sanders_24} calibration, similar to \citetalias{Chakraborty_24} calibration. Aside from the offset in absolute normalization, our observed MZR is consistent with \citet{Sarkar_25,Nakajima_23}, while \citet{Curti_24} reports a flatter slope. Compared with observations at lower redshift, the MZR in the same stellar mass range shows an offset of $\sim0.6$ dex lower than \citet{He_24} at $z\sim3$. Together with \citet{Nakajima_23,Sarkar_25,He_24}, for low mass galaxies ($\log(M_*/M_\odot)<9$) the MZR slope at $2<z<10$ is slightly shallower but not significantly different from that of more massive galaxies ($\log(M_*/M_\odot)>9$) at $z\sim2-3$ as reported by \citet{Sanders_21} from the MOSDEF survey. While \citet{Curti_24,Lim_23} presented a shallower MZR slope ($\gamma\sim0.16-0.17$ at $2<z<9$) , and \citet{Heintz_23,Chemerynska_24} show steeper MZR slopes ($\gamma\sim0.33-0.39$ at $z>6$).

\subsubsection{MZR of field and cluster galaxies}

\begin{figure}[htbp!]
    \centering
    \resizebox{9cm}{!}{\includegraphics{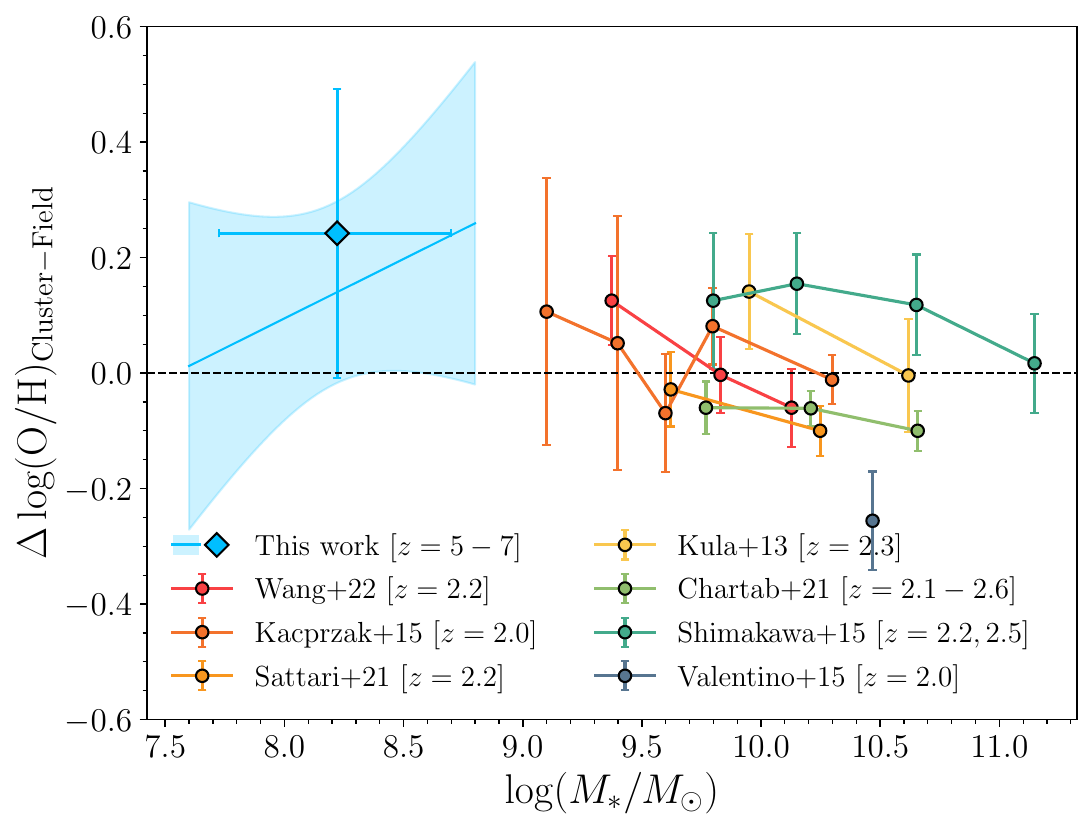}}
    \caption{The metallicity offset between galaxies in overdense and field environments. The blue line shows the offset of MZR between cluster and field galaxies, with the shaded region representing $1\sigma$ confidence interval. The single blue diamond shows the offset between the mass-matched cluster and field samples. The measurements in literature \citep{Wang_22,Kacprzak_15,Sattari_21,Kulas_13,Chartab_21,Shimakawa_15,Valentino_15} are also shown for comparison, each marked in different colors. The error bars represent $1\sigma$ uncertainties.}\label{fig:d_oh}
    \label{fig:d_OH}
\end{figure}

We perform the same linear regression on the field and cluster stacks in our sample. 
In Fig. \ref{fig:mzr}, we show the MZR measured for field and cluster galaxies using \citetalias{Chakraborty_24} calibration, and in Fig. \ref{fig:d_oh} we show the difference between field and cluster MZRs in comparison with literature studies.
We find that galaxies at $5<z<7$ in overdense environments exhibit a steeper MZR slope than coeval field galaxies, and cluster galaxies are more metal-rich than field galaxies, especially at the high-mass end by $\sim0.2-0.3$ dex, indicating enhanced metal-enrichment processes in overdense environments.
As a comparison, we also measure the MZR using \citetalias{Sanders_24} calibration, and the results are listed in Table \ref{tab:mzr_info}. The \citetalias{Sanders_24} calibration gives a shallower slope than that using \citetalias{Chakraborty_24} calibration. Thus, we caution that the choice of calibration can affect the observed MZR slope. Despite the systematic difference in different calibrations, the steeper MZR slope of cluster galaxies than field galaxies is confirmed regardless of calibrations. However, we note that, given the typical uncertainties of $\gtrsim0.1$ dex in metallicity measurements, the observed metal enhancement in overdensities is mild and remains compatible within the measurement errors.

We also observe that the galaxies in overdense environments are more massive than field galaxies at the same redshift, as in each mass bin, the median stellar mass of cluster galaxies is $\sim0.1-0.2$ dex higher than that of field galaxies. More massive galaxies are intrinsically more metal-rich.
Although this is expected to be a small effect, to reduce the potential bias of different stellar mass distributions in field and cluster galaxies, we additionally build a mass-matched field sample. We select the field galaxy that is closest in stellar mass to each cluster galaxy and measure the metallicity in the same way discussed above. 
The measurements in the three mass bins of the cluster and field subsets are listed in Table \ref{tab:r3_info}. We find that, at the same stellar mass, the cluster galaxies are overall metal-richer by $\sim0.2$ dex compared to their field counterparts, regardless of the choice of \citetalias{Sanders_24} or \citetalias{Chakraborty_24} R3 calibrations. The offset of the mass-matched sample is also presented in Fig.~\ref{fig:d_oh}, and remains consistent with the predicted offsets in MZR.

\section{The fundamental metallicity relation}\label{sec:FMR}

The fundamental metallicity relation (FMR) relates metallicity, stellar mass and star formation rate (SFR) \citep[e.g.,][]{Lara-lopez_10,Mannucci_10,Curti_20,Sanders_21,Baker_23}. It is a key scaling relation to understand the physical processes in galaxy evolution. \citet{Ellison_08} show that galaxies with higher SFR systematically exhibit lower metallicity than those with lower SFR at fixed stellar masses. \citet{Lara-lopez_10} and \citet{Mannucci_10} further show that the scatter in the mass-metallicity relation can be reduced if the SFR is introduced as a secondary parameter with stellar mass, i.e., in terms of the SFR--MZ relation.
We apply the parametrization of \citet{Andrews_13} to describe the SFR--MZ relation by finding the value of $\alpha$ that minimizes the scatter in the metallicity at a fixed $\mu_\alpha$, defined as:

\begin{equation}
    \mu_{\alpha} = \log \left( \frac{M_*}{M_{\odot}} \right) - \alpha \log \left( \frac{\text{SFR}}{M_{\odot} \text{ yr}^{-1}} \right).\label{eq:mu_alpha}
\end{equation}
We estimate the SFR by using \Hb\ luminosity and the \cite{Kennicutt_98} calibration corrected to a Kroupa IMF \citep[as applied by][]{Heintz_23,Sarkar_25}:

\begin{equation}
    \mathrm{SFR}=5.5\times10^{-42}\frac{L_{\Hb}}{\rm erg\ s^{-1}} (M_\odot\ \rm yr^{-1})\times f_{\Ha/\Hb},
\end{equation}
where we convert their calibration using the theoretical ratio $f_{\Ha/\Hb}=2.86$ with case B recombination \citep{Osterbrock_89}. 
The $L_{\Hb}$ is measured from the median stacked spectra. Since $\Hg$ is partially covered in our sample at $z>6.25$, we are not able to estimate dust attenuation for the full sample. From our $z>6.25$ stacks, the dust attenuation is estimated as $A_V=0.38^{+0.54}_{-0.38}$, and we use the same dust attenuation for all the sample galaxies at $5<z<7$. Considering the large uncertainty of $A_V$, we do not directly correct for dust on $L_{\Hb}$ measurements; instead, we use $A_V=0.38$ as an upper bound for the $L_{\Hb}$ error, corresponding to $\sim50\%$ flux uncertainty. We add the $\sim50\%$ uncertainty into the statistical error in quadrature and propagate it into the SFR estimation. 

The estimated SFRs from the median stacks are $\left[11.78^{+5.90}_{-0.34},13.54^{+6.79}_{-0.46}, 8.49^{+4.28}_{-0.50}\right]~M_\odot/\mathrm{yr}$, for the $[\rm full,field,cluster]$ samples, respectively. For comparison, we also report the median values of the SED-based SFRs: $\left[9.20, 10.14, 8.74\right]~M_\odot/\mathrm{yr}$. The SFRs derived from \Hb\ are generally higher than those estimated from SED fitting, as the Balmer line \Hb\ reflects the recent star formation, while the SFR derived from SED fitting traces the average star formation rate over a longer timescale.

We follow \citet{Andrews_13} who found that $\alpha=0.66$ minimizes the scatter in $\mu_\alpha-\text{metallicity}$ plane, for a sample of local low mass galaxies down to $M_*=10^{7.4}M_\odot$ with direct $T_e$ metallicity. They showed that with $\alpha=0.66$, the $\mu_\alpha-\text{metallicity}$ relation is parameterized as:
\begin{equation}
    12 + \log(\mathrm{O/H}) = 0.43\times\mu_{0.66}+4.58.\label{eq:fmr}
\end{equation}

We note that there are some discrepancies in the literature about the value of $\alpha$. With SDSS spectra, \citet{Curti_20} and \citet{Sanders_21} suggested $\alpha=0.55$ and $\alpha=0.60$ respectively, while several studies suggest weaker dependence on SFR, i.e., $\alpha\lesssim0.4$ \citep{Mannucci_10,Guo_16,Henry_21}. 
The discrepancy in $\alpha$ can be attributed to the sample selections and the choice of the metallicity calibrations. A more detailed discussion about the value of $\alpha$ is beyond the scope of this paper. 
We use $\alpha=0.66$ as is widely used for comparison with recent JWST observations at similar redshifts and stellar mass range \citep{Nakajima_23,Sarkar_25}.

Previous studies have shown varying results regarding the offset from the local FMR. \citet{Curti_24} and \citet{Heintz_23} reported a significant offset starting from $z\sim4$, while \citet{Nakajima_23} and \citet{Sarkar_25} found no significant deviation at $z<7$. The differences in these findings can be attributed to the metallicity calibrations used. \citet{Nakajima_23} and \citet{Sarkar_25} employed calibrations from \citetalias{Nakajima_22} and \citetalias{Curti_20}, which may overestimate the metallicities with respect to \citetalias{Chakraborty_24} by about 0.2 dex, as noted in \citet{Chakraborty_24}. \citet{Heintz_23} utilized the \citetalias{Sanders_24} calibration, which is based on high-redshift direct $T_e$ measurements, making it more consistent with our findings.

Additionally, the choice of $\alpha$ in the FMR formalism affects the results. Comparing the metallicities from references adopting different alpha values could result in a misinterpretation of the predicted metallicities.
\citet{Heintz_23} used $\alpha = 0.55$ from \citet{Curti_20}, leading to a metallicity prediction that is $\sim0.1$ dex higher than predictions using $\alpha = 0.66$ from \citet{Andrews_13}. \citet{Nakajima_23} also found that using the FMR formalism of \citet{Curti_20} predicts a higher metallicity than using the formalism of \citet{Andrews_13} at $z>6$, and thus get a more negative offset. 
This difference in $\alpha$ leads to a larger offset observed by \citet{Heintz_23} compared to \citet{Sarkar_25} and \citet{Nakajima_23}. While we expect a smaller offset for \citet{Heintz_23} if $\alpha = 0.66$ is used, there is still a significant offset from the local FMR.
Furthermore, \citet{Curti_24} employed a different FMR formalism from \citet{Curti_20}, yet this is similar to the parametrization using $\alpha = 0.55$. 

On the other hand, different SFR conversions are applied in different works. \citet{Nakajima_23} assumed Chabrier IMF when converting the \Ha\ or \Hb\ luminosity to SFR, while \citet{Heintz_23} and \citet{Sarkar_25} assumed the Kroupa IMF. 
The main FMR results reported by \citet{Curti_20} are based on their SED-derived SFRs. They also provided SFRs converted from \Ha\ luminosities, using a calibration derived from low-metallicity galaxies in \citet{Reddy_22}. The \citet{Reddy_22} calibration yields lower SFRs that are about $40\%$ of those obtained using a Kroupa IMF. In contrast, the Kroupa IMF yields SFRs that are higher than those from the Chabrier IMF by $\sim 6\%$ \citep{2014ARA&A..52..415M}, and has a negligible impact on the FMR \citep{Nakajima_23}.

Considering the difference in calibrations and FMR formalism used in different studies, we have recalculated the FMR offsets based on \citet{Andrews_13} formalism with $\alpha = 0.66$ for a fair comparison in this work. Since the difference between Kroupa and Chabrier IMFs is negligible, we do not apply any conversion between them, as used by \citet{Nakajima_23,Heintz_23,Sarkar_25}. However, we convert the \citet{Reddy_22} calibration to a Kroupa IMF for consistency.
We note that the recalculated values remain consistent with the original measurements within the quoted uncertainties.
Below we list the details of their measurements and our recalculations.
\begin{enumerate}
    \setlength{\itemsep}{0.5em}
    \item \citet{Curti_24} used a revised version of \citetalias{Curti_20} calibration that is refined by a sample of local metal-poor galaxies. We retrieved their measurements from Table B.1 and recalculated the FMR offset using \citet{Andrews_13} formalism. To account for systematic differences between local and high-redshift $T_e$ measurements noted by \citet{Chakraborty_24}, we applied a $-0.2$ dex offset to their metallicities. This offset is based on empirical differences between calibrations (see Fig. 7 in \citealt{Chakraborty_24}). A more rigorous approach would require recalculating from original line flux measurements, which is beyond our scope. Their SFRs derived from \Ha\ or \Hb\ fluxes are based on \citet{Reddy_22} calibration, and we converted the SFRs to Kroupa IMF by multiplying a factor of 2.6.
    \item \citet{Heintz_23} used \citetalias{Sanders_24} calibration, which is calibrated using high-redshift direct $T_e$ measurements, and is more consistent with \citetalias{Chakraborty_24} calibration. We do not apply an additional metallicity offset for their measurements. Their offset from FMR is calculated using $\alpha = 0.55$ from \citet{Curti_20}. We retrieved their measurements from their Table 1 and recalculated the offset following \citet{Andrews_13}. 
    \item \citet{Nakajima_23} used \citetalias{Nakajima_22} calibration, which is based on a sample of local metal-poor galaxies. They used the same FMR formalism as \citet{Andrews_13}. We added a constant offset of $-0.2$ dex to their FMR offsets to account for the systematic offset in metallicity calibrations. 
    \item \citet{Sarkar_25} used \citetalias{Curti_20} calibration. They used the same FMR formalism as \citet{Andrews_13}. We added a constant offset of $-0.2$ dex to their FMR offsets to account for the systematic offset in metallicity calibrations. 
\end{enumerate}

\begin{figure}[t!]
    \centering
    \resizebox{8.5cm}{!}{\includegraphics{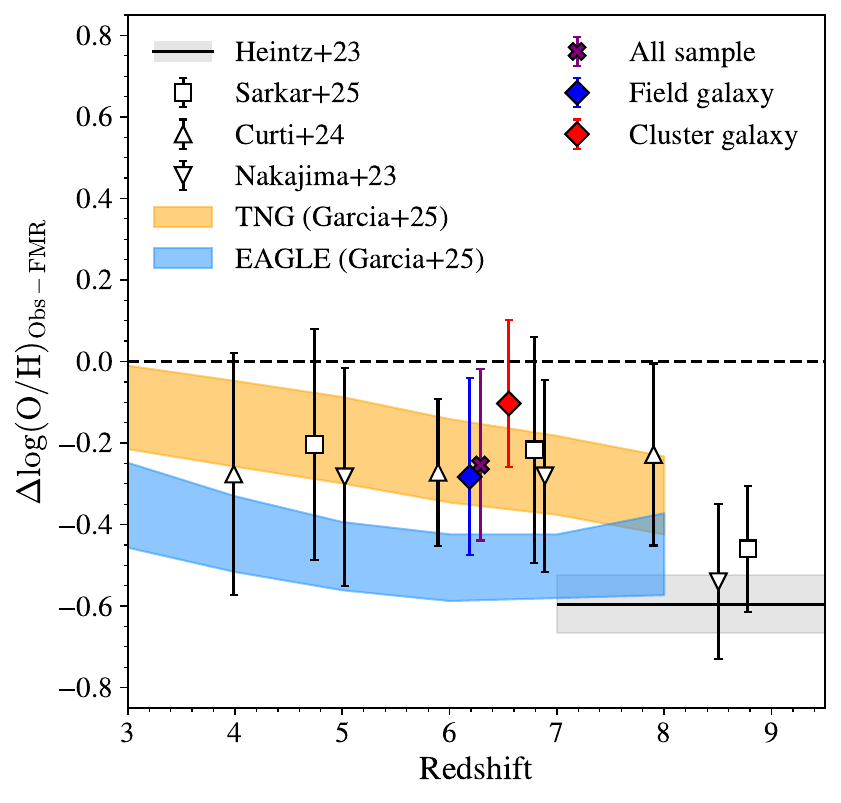}}
    \caption{The offsets in metallicity between observations and the predictions of the local FMR with $\alpha = 0.66$ \citep{Andrews_13}. The orange and blue shadows represent the predictions in TNG and EAGLE simulation \citep{Garcia_25}. The offsets reported by \citet{Curti_24,Heintz_23} used a different FMR formalism, and we recalculated the offsets using \citet{Andrews_13} formalism. For \citet{Nakajima_23,Sarkar_25}, we added a constant offset of $-0.2$ dex to the their FMR offsets to account for the systematic offset in metallicity calibrations used in their studies. 
    }
    \label{fig:fmr}
\end{figure}

In Fig. \ref{fig:fmr}, we compare the observations of our sample galaxies with the predictions of the local FMR. We find that the full sample galaxies are approximately 0.25 dex lower than the local FMR predictions. Cluster galaxies, however, are more metal-rich and slightly offset from the local FMR. From the recalculated offsets in literature studies, we find a similar metal dilution at $z>5$ as our observations. \citet{Heintz_23} interprets such an offset by the effective dilution due to pristine gas infall onto the galaxy. In hydrodynamical simulations, \citet{Garcia_24} found a non-negligible evolution in $\alpha$ as a function of redshift, rather than assuming a constant value as we have done with $\alpha = 0.66$. \citet{Garcia_25} further suggested that the role SFR plays in setting the normalization may change with redshift, and an increasing anti-correlation between MZR and SFR possibly leads to the observed metal deficiency compared with FMR observed at $z=0$. The evolving FMR may indicate changing physical processes influencing galaxy metal enrichment, such as gas accretion and feedback.

\section{Simple analytical model of chemical evolution}\label{sec:model}
\subsection{Model description}
To help understand the physical origin of star formation and metal enrichment, many analytical models of galaxy chemical evolution have been proposed \citep{Edmunds_90,Edmunds_95,Koppen_99,Finlator_08,Erb_08,Dave_12,Dayal_13,Lilly_13,Peng_14,Guo_16,Wangen_21,Toyouchi_25}. 
Several studies have used a ``closed box'' model or ``leaky box'' model to explain the evolution of MZR \citep{Ma_16,Langan_20}, which is a modified version of ``closed box'' model, while using an effective yield ($p_\text{eff}$) instead of the yield ($p$) to account for flows of metals through inflow and outflow. The yield $p$ is defined as the mass in metals returned to the ISM relative to the mass in stars formed: $p=\dot{M}_{Z,~\text{ejected}}/\text{SFR}$ \citep{Pagel_75}.
In the ``leaky box'' model, the ISM metallicity is simply a function of effective yield and gas fraction:
\begin{equation}
    Z_\text{ISM}=-p_\text{eff}\cdot\ln\left(f_\text{gas}\right),\label{eq:Z_leaky_box}
\end{equation}
where $f_{\text{gas}}$ is defined as the mass fraction of gas to the total mass of gas and stellar components.
When $p_\text{eff}=p$, the system is a ``closed box'' and the metallicity only depends on the consumption of the initial gas reservoir and metal production within the system. When $p_\text{eff}<p$, the system is ``leaky'' and the metals are lost by metal-enriched outflow. However, the detailed pathways of those metal flows are not explicitly modeled.

To capture more details, here we model the galaxy ISM as an open system with gas inflow and outflow, and consumption of gas by star formation (also known as the ``bathtub'' model) \citep[See][for helpful reviews]{Pagel_09,DynamicsAstrophysicsGalaxies}. Following \citet{Erb_08}, with the conservation of gas mass, we have the relation:
\begin{equation}
    \dot{M}_\text{gas}=\dot{M}_{\text{in}}-\dot{M}_{\text{out}}-\mu\text{SFR},\label{eq:gas_balance}
\end{equation}
where $\dot{M}_{\text{in}}$ and $\dot{M}_{\text{out}}$ are the mass flow rate of gas inflow and outflow, and $\mu$ is the fraction of mass that is locked in long-lived stars and remnants, i.e., $(1-\mu)$ represents the efficiency of gas return from SNe. 

With the conservation of metal mass, we have:
\begin{equation}
    \begin{aligned}
        \dot{M}_Z=\dot{M}_{\text{in}}Z_{\text{in}}+y\mu (1-Z_\text{SFR}) \text{SFR}-Z_{\text{out}}\dot{M}_{\text{out}}-\mu\text{SFR}\cdot Z_\text{SFR},\label{eq:metal_balance}
    \end{aligned}
\end{equation}
where $Z_{\text{in}}$ is the metallicity of gas inflow, $Z_{\text{out}}$ is the metallicity of gas outflow, and $Z_\text{SFR}$ is the metallicity of the gas forming stars, and $y$ is the metal yield.
The first term on the right-hand side is the metal mass flow rate from the gas inflow. The second term represents the metal enrichment from metals produced from star formation.
We note that we use a different definition of yield $y$ \citep[e.g., applied by][]{Erb_08}, to be distinguished from $p$ in Eq. \ref{eq:Z_leaky_box}. Here, $y$ is defined as the mass in
metals returned to the ISM relative to the mass of hydrogen consumed by star formation, instead of total stellar mass formed: $y=\dot{M}_{Z,~\text{ejected}}/\dot{M}_\text{H,~consumed}$ \citep{Searle_72}.
With this definition, $(1-Z_\text{SFR})$ represents the fraction of hydrogen that contributes to chemical enrichment, given that the metals incorporated into star formation are not further available for subsequent metal production. Other works using the definition $p$ do not include the $(1-Z_\text{SFR})$ term \citep[e.g.,][]{Peng_14}.
Since the metal content is low relative to hydrogen, these two different definitions do not result in any significant differences in reasonably low metallicity ranges.
The third term is the metal mass loss from gas outflow, and the fourth term is the metal consumption when ISM gas, together with metals therein, collapses into stars. 

For simplicity, here we assume that $\mu=1$, i.e., gas returned to the ISM is negligible.
We assume that the metallicity of gas forming stars is the same as ISM gas: $Z_\text{ISM}=Z_\text{SFR}$, and the inflowing gas is metal-free: $Z_{\text{in}}=0$. The outflow metallicity is assumed to be the same as the ISM metallicity.
We further assume a steady-state approximation: the gas and metal masses are in the state of dynamic equilibrium: $\dot{M}_\text{gas}=0,~\dot{M}_\text{Z}=0$ \citep[also see][]{Dave_12}. 
For low metallicity galaxies with $Z_\text{SFR}\ll1$ such that $1-Z_\text{SFR}\approx1$, Eq. \ref{eq:metal_balance} indicates the ISM metallicity:
\begin{equation}
    Z_\text{ISM}=\frac{y\cdot\text{SFR}}{\dot{M}_{\text{out}}+\text{SFR}}.\label{eq:Z_ISM}
\end{equation}

The mass outflow rate is related to the SFR and mass loading as: 
\begin{equation}
    \dot{M}_{\text{out}}=\eta\cdot \text{SFR}^{\lambda},
\end{equation}
where $\eta$ is the mass loading factor, defined as the amount of gas ejected per unit mass of star formation, and $\lambda>0$ allows the outflow rate to non-linearly correlate with SFR. E.g., simulations suggest that if stars form in clusters with a higher SFR, the successive feedback from supernovae (SNe) in clusters can create superbubbles large enough to break out of the galactic disk, enhancing the outflow efficiency \citep{Orr_22}.
Wind models show that the mass loading factor is related to halo masses: $\eta\propto M_{\text{h}}^{-1/3}$ for momentum-driven wind (kinetic feedback) and $\eta\propto M_{\text{h}}^{-2/3}$ for energy-driven wind (mechanical feedback), derived either via momentum or energy conservation \citep{Dave_12}. 
The halo mass can be estimated with Stellar-to-Halo mass relation: $M_h=\mathcal{F}(M_*)$ \citep[e.g.,][]{Moster_10,Behroozi_13,Shuntov_22,Shuntov_24}. We have:
\begin{equation}
    \eta=\eta_0\cdot\left(\frac{\mathcal{F}(M_*)}{M_0}\right)^{-\beta},\label{eq:eta}
\end{equation}
where $\eta_0$ is the mass loading factor at a reference halo mass $M_0$, and $\beta$ characterizes the mass dependence of the mass loading factor, i.e., $\beta=1/3$ for momentum-driven wind and $\beta=2/3$ for energy-driven wind. Different $\beta$ values are also adopted in other semi-analytical models to treat SNe feedback \citep{Cole_94,Guo_11}. We do not include AGN feedback in our model, as it is not the primary focus for young star-forming galaxies.
Replacing $\dot{M}_{\text{out}}$ in Eq. \ref{eq:Z_ISM}, we have:
\begin{equation}
    Z_\text{ISM}=\frac{y}{1+\eta \text{SFR}^{\lambda-1}}.\label{eq:Z_steady}
\end{equation}

The above solution relies on a steady-state approximation. However, if the galaxy is accumulating gas more rapidly than its depletion, i.e., $\dot{M}_{\text{gas}}>0$, the galaxies live in a non-equilibrium state \citep[e.g.,][]{Peng_14}. We could perturb the above steady-state solution by adding an instant gas dilution term:
\begin{equation}
    \begin{aligned}
    Z_\text{ISM,obs}&=Z_\text{ISM,steady}/\left(1+\frac{\dot{M}_{\text{in}}(\tau_\text{depl}-\tau_\text{acc}) }{M_\text{gas}}\right)\\
    &=Z_\text{ISM,steady}\cdot\tau_\text{acc}/\tau_\text{depl},
    \end{aligned}\label{eq:Z_obs_dilution}
\end{equation}
where $\tau_\text{acc}$ is the gas accretion time-scale defined as $\tau_\text{acc}=M_\text{gas}/\dot{M}_{\text{in}}$, and $\tau_\text{depl}$ is the gas depletion time-scale: $\tau_\text{depl}=M_\text{gas}/\text{SFR}$. This describes the amount of extra gas accreted onto the galaxy in an episode of gas depletion after prior gas accretion. If $\tau_\text{depl}>\tau_\text{acc}$, gas accretion is faster than gas depletion, and the ISM metallicity is more diluted than its steady-state solution.
We define $\epsilon_\text{acc}$ as the efficiency relating SFR to the gas accretion rate: 
\begin{equation}
    \epsilon_\text{acc}=\text{SFR}/\dot{M}_{\text{in}},\label{eq:epsilon_acc}
\end{equation}
which describes the fraction of accreted gas that is converted to stars \citep[e.g.,][]{Dave_12}. Similarly, we also define the star formation efficiency\footnote{This definition should not be confused with the cosmological SFE, which is the galaxy stellar mass divided by the halo baryon mass $M_*/(f_\text{b}M_\text{halo})$, where $f_\text{b}$ is the baryon fraction \citep[e.g.,][]{Dekel_23}.} as the ratio of SFR to the total gas mass: 
\begin{equation}
    \epsilon_\text{SF}=\text{SFR}/M_\text{gas}.\label{eq:epsilon_SF}
\end{equation} 
The Kennicutt-Schmidt law \citep[KS law,][]{Schmidt_59,Kennicutt_89} describes the relation between surface densities of star formation and gas mass: 
\begin{equation}
    \Sigma_\text{SFR}\propto\Sigma_\text{gas}^{n} \label{eq:KS_law}
\end{equation}
\citet{Kennicutt_98} suggested that $n\approx1.4$ for local star-forming galaxies. 
The total star formation rate is the integral of the surface density of star formation over the surface area ($s$) of the galaxy:
\begin{equation}
    \text{SFR}\propto\int_s\Sigma_\text{gas}^{n}ds.
\end{equation}
With a crude approximation that the gas surface density is a constant across the galaxy, and is not evolving with time, even though the gas mass is changing, the SFR is linearly correlated with the gas mass as one galaxy evolves:
\begin{equation}
    \text{SFR}(t)\propto M_\text{gas}(t)\rightarrow \text{SFR}(t)=\epsilon_\text{SF}M_\text{gas}(t).
\end{equation}
SFR no longer explicitly depends on $n$, naturally following from Eq. \ref{eq:epsilon_SF}. In this sense, the galaxy is accumulating mass with growing size/volume, while keeping the same gas density, although this might not hold as the gravitational potential changes. Here, the $\epsilon_\text{SF}$ implicitly describes the KS law, with a larger $\epsilon_\text{SF}$ indicating a higher gas density, which produces a higher SFR. In our assumption, the $\epsilon_\text{SF}$ is a constant through the evolution of a single galaxy, but the value can vary for different galaxies with different gas densities to keep the KS law.

The above relations naturally suggest $\tau_\text{acc}/\tau_\text{depl}=\epsilon_\text{acc}$, so that the observed metallicity from Eq. \ref{eq:Z_obs_dilution} is then:
\begin{equation}
    Z_\text{ISM,obs}=Z_\text{ISM,steady}\cdot\epsilon_\text{acc},
\end{equation}
and in log scale:
\begin{equation}
    \begin{aligned}
    \log Z_\text{ISM,obs}&=\log Z_\text{ISM,steady}+\log\epsilon_\text{acc}\\
    &=\log y-\log\left(1+\eta \text{SFR}^{\lambda-1}\right)+\log\epsilon_\text{acc}.
    \end{aligned}
\end{equation}
If the outflow rate is dominant, i.e., $\eta \text{SFR}^{\lambda-1}\gg1$, the observed metallicity is then approximated as:
\begin{equation}
    \log Z_\text{ISM,obs}\approx\log y-\log\eta+\log\epsilon_\text{acc}-(\lambda-1)\log\text{SFR}. \label{eq:Z_obs}
\end{equation}
This solution relates the observed metallicity to the steady-state metallicity, the SFR, and the stellar mass. With a fixed stellar mass and gas mass, the mass loading term is a constant. The observed metallicity is then a simple linear function of $\log\text{SFR}$:
\begin{equation}
    \log Z_\text{ISM,obs}\propto-(\lambda-1)\log\text{SFR}.
\end{equation}
This is in the same form as the formalism of FMR in observations, e.g., Eqs. \ref{eq:mu_alpha}, \ref{eq:fmr}.

In cases where outflow is small, i.e., $\eta \text{SFR}^{\lambda-1}\ll1$, the observed metallicity is then approximated as:
\begin{equation}
    \log Z_\text{ISM,obs}\propto-\eta\text{SFR}^{\lambda-1}.
\end{equation}
This relation is less linear, but also predicts the decreasing metallicity with increasing SFR.

We note that the observed metallicity actually varies with the age of the galaxy we observe.
A more strict solution can be derived by solving the differential equations using Eq. \ref{eq:gas_balance} and Eq. \ref{eq:metal_balance} and inferring the metallicity evolution with galaxy ages. The strict solution also suggests that all galaxies will eventually reach their steady-state metallicities at sufficiently large age, no matter how much the gas accretion rate is. We also note that $\epsilon_\text{acc}$ is actually not a constant, and varies with SFR as galaxies evolve. The dilution by the factor of $\epsilon_\text{acc}$ is a simple proxy for the delay of metal enrichment due to intense gas accretion at the beginning of galaxy formation. As a galaxy evolves, the gas reservoir builds up and reaches equilibrium, the star formation rate peaks with a higher $\epsilon_\text{acc}$, and the dilution effect is reduced. 

To better illustrate such time evolution during a non-equilibrium state, we further discuss the results of numerical solutions in the context of MZR and FMR evolution in Section \ref{sec:numerical_solutions} and Section \ref{sec:fmr_evolution}.

\subsection{Numerical solutions of model equations}\label{sec:numerical_solutions}
\citet{Mannucci_10} suggested that the FMR originates when galaxies are observed in a transient phase where the dilution effect from gas infall dominates over the metallicity enrichment from newly formed stars; thus, understanding the time evolution of the mass assembly and metal enrichment can be important.
We show that this can be revealed from the solutions of Eq. \ref{eq:gas_balance} and Eq. \ref{eq:metal_balance}. We have constructed different sets of models exploring different parameter spaces. In all models with outflow, we have assumed feedback is linearly correlated with SFR, i.e., $\lambda=1$. We use the same initial conditions for all the models with inflow: a pre-existing gas reservoir $M_\text{gas,0}=10^{6}M_\odot$, with few stars $M_{*,0}=10^{1}M_\odot$ and metal-free $Z_\text{ISM,0}=0$. While for models without inflow, we assign different pre-existing gas mass $10^{5.0}-10^{10.0} ~M_\odot$ for different galaxies.
Following \citet{Erb_08}, we use the stellar yield $y=0.019=1.5Z_\odot$ for all the models, where $Z_\odot=0.0126$ is the solar metallicity \citep{Asplund_04,Asplund_09}. In the steady-state solution, we have assumed $\mu=1$ as no gas recycling in Eq. \ref{eq:gas_balance}. However, the gas returned from massive stars can be non-negligible for a longer duration of star formation, e.g., the fraction of mass returned from stars approaches $40\%$ of the total stellar mass formed for a Chabrier IMF in a few Gyr \citep{Erb_08}. For a top-heavy IMF, the fraction of mass returned can be higher. While the time-varying return fraction can be non-trivial, we approximate a time-invariant return fraction $40\%$ with $\mu=0.6$ for simplicity. This implies the instantaneous recycling approximation, which is generally appropriate for metal yields from short-time-scale massive-star evolution, particularly oxygen \citep{Pagel_09}. We vary other parameters to explore the effects of different feedback strengths and star formation efficiencies in the following five model sets.

\begin{itemize}
    \setlength{\itemsep}{0.5em}
    \item Model 1: Incorporating inflow but no outflow, with a varying star formation efficiency $\epsilon_\text{SF}=[0.02,0.05,0.1] \times 10^{6} ~\text{Myr}^{-1}$. Observations of giant molecular clouds in the Milky Way suggest a star formation efficiency per free-fall time in the range of $0.002$–$0.2$ \citep{Murray_11}. Our adopted values are therefore reasonable, given that the free-fall time for galaxies at $z > 5$ could be on the order of $\sim$1 Myr \citep{Dekel_23}.
    We fix $\eta_0=0$ so that no gas is removed from the galaxy. We use the constant gas accretion rate between $10^{5.5}-10^{9.0} ~\text{M}_\odot~\text{Myr}^{-1}$ for different galaxies.
    \item Model 2: Incorporating outflow but no inflow, varying star formation efficiency $\epsilon_\text{SF}=[0.02,0.05,0.1] \times 10^{6} ~\text{Myr}^{-1}$. We fix $\dot{M}_\text{in}=0$ so that no gas is accreted from the intergalactic medium, and the galaxy only consumes the gas pre-existing in the reservoir, set as $10^{5.0}-10^{10.0} ~M_\odot$ for different galaxies.
    \item Model 3: Incorporating both inflow and outflow, varying mass loading factor $\eta_0=[0.5,1,2]$. We assume the same normalization halo mass $M_0=10^{10}M_\odot$ in Eq. \ref{eq:eta}. We apply the Stellar-to-Halo mass relation at $z=5$ from \citet{Shuntov_22} to convert the stellar mass to the halo mass, which constrains the outflow rate in Eq. \ref{eq:eta}. We use the constant gas accretion rate between $10^{5.5}-10^{9.0} ~\text{M}_\odot~\text{Myr}^{-1}$ for different galaxies.
    \item Model 4: Incorporating both inflow and outflow, varying outflow mode $\beta=[0,1/3,2/3,1]$.  Specifically, $\beta=0$ corresponds to the constant wind model ($\eta=\eta_0$), $\beta=1/3$ corresponds to the momentum-driven wind model, and $\beta=2/3$ corresponds to the energy-driven wind model. In addition, $\beta=1$ corresponds to a stronger wind model, where the outflow rate scales more significantly with halo mass. The physical mechanisms of SNe feedback remain less clear when assuming $\beta=1$. However, similar values are also adopted in semi-analytical models to reproduce the observed luminosity or stellar mass functions \citep[e.g.,][]{Guo_11}. Other parameters in this set are the same as in Model 3. We use the constant gas accretion rate between $10^{5.5}-10^{9.0} ~\text{M}_\odot~\text{Myr}^{-1}$ for different galaxies.
    \item Model 5: Incorporating both inflow and outflow, with parameters fixed to match the observations at $5<z<7$. We use $\epsilon_\text{SF}=0.02\times10^{6} ~\text{Myr}^{-1}$, $\eta_0=5$ and $\beta=2/3$.
\end{itemize}

\begin{figure*}[t!]
    \centering
    \resizebox{18cm}{!}{\includegraphics{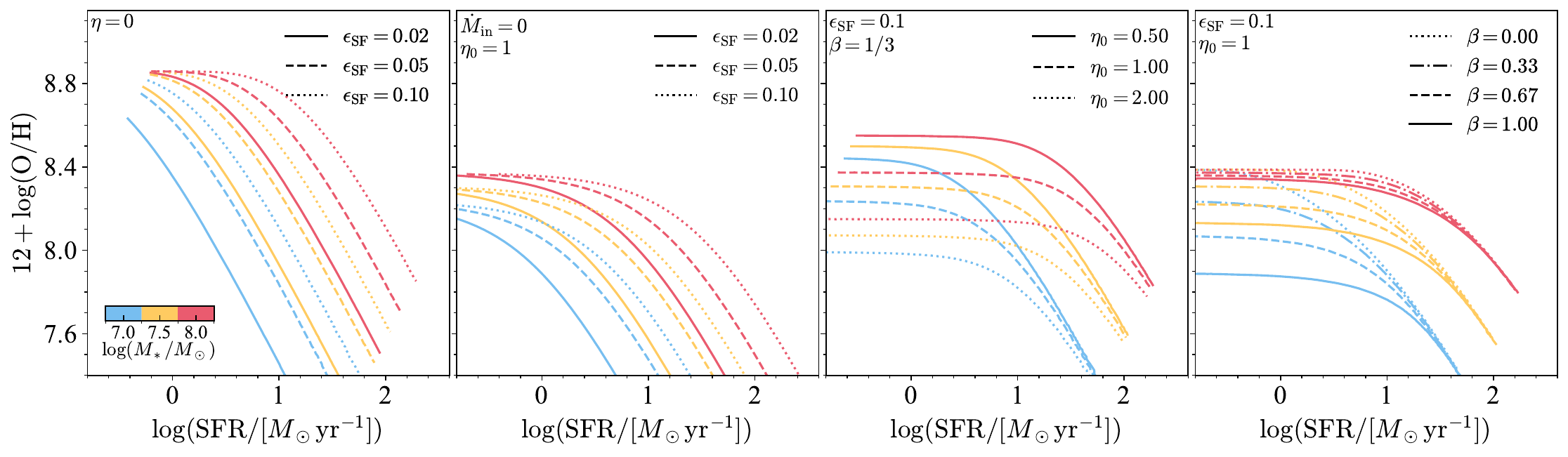}}
    \caption{The predicted $Z-\text{SFR}$ relations for different model configurations. The first panel: Model 1, incorporating inflow but no outflow, with $\epsilon_\text{SF}$ as the varying parameter. The second panel: Model 2, incorporating outflow but no inflow, with $\epsilon_\text{SF}$ as the varying parameter. The third panel: Model 3, incorporating both inflow and outflow, with $\eta$ as the varying parameter. The fourth panel: Model 4, incorporating both inflow and outflow, with $\beta$ as the varying parameter. In all panels, the lines are color-coded by stellar mass, while different line styles indicate variations in model parameters.}
    \label{fig:FMR_predictions}
\end{figure*}

We obtain the solutions of the above models by integrating the differential equations using $4^\text{th}$ order Runge-Kutta method. The results contain arrays of time and other variables at each time step. To explore the FMR evolution, we match SFR and stellar mass for different galaxies. The results are shown in Fig. \ref{fig:FMR_predictions}. Similarly, to obtain the time evolution of MZR, we define the $t_\text{obs}$ of a galaxy as the time elapsed since the onset of gas accretion and star formation, and we pick up galaxies with different $t_\text{obs}$. Using model 5, we show the predicted MZRs with different $t_\text{obs}$ in Fig. \ref{fig:MZR_predictions}. 
In Fig. \ref{fig:FMR_predictions}, we see that the high-mass galaxies show a flatter $Z-\text{SFR}$ relation than low-mass galaxies, and at fixed stellar masses, a higher star formation rate leads to lower metallicities, which are qualitatively consistent with the observations \citep[e.g.,][]{Andrews_13,Curti_20}. 
We discuss the implications of the results in the following Section \ref{sec:fmr_evolution}.

\subsection{Model implications of FMR evolution}\label{sec:fmr_evolution}

For the case where FMR is independent of redshift, such a non-evolving FMR implies that the metallicity is a fixed function of SFR, at fixed stellar masses. Under steady-state condition, this indicates that Eq. \ref{eq:Z_obs} is not changing with redshift, which requires that:
\begin{equation}
    \left\{
    \begin{aligned}
        &\frac{d}{dt}\log\left(\frac{y\epsilon_\text{acc}}{\eta}\right)=0,\\
        &\frac{d}{dt}\lambda=0.
    \end{aligned}
    \right.
\end{equation}
A constant $\lambda$ indicates that the outflow gas has the same response to star-forming feedback. And a constant $y\epsilon_\text{acc}/\eta$ requires that the metal yield, gas dilution, enrichment, and mass loading factor are either constant or maintain the same balance across cosmic time. 

Before $z\sim3$, observations show that the FMR is nearly independent of redshift \citep{Mannucci_10,Henry_21}. However, literature studies suggest that the FMR starts to evolve at $z\gtrsim3$ \citep{Mannucci_10,Curti_24}. \citet{Moller_13} suggests a transition phase of primordial gas infall at $z\sim2.6$, which may be relevant for such evolution.

We now explore the possible consequences of FMR evolution in response to changes in factors within our simple model.

\subsubsection{Gas accretion and star formation efficiency}
Eq. \ref{eq:Z_obs_dilution} indicates that a lower metallicity is related to a lower efficiency $\epsilon_\text{SF}$, which happens when the gas accretion rate is high, and the metal enrichment of ISM is delayed. In this case, the gas reservoirs of galaxies at $z>5$ are in a more non-equilibrium state, with high gas fraction and low metallicity. This is consistent with observations of abundant gas reservoirs at high redshifts \citep{Heintz_22}. \citet{Dekel_23} proposed the Feedback-Free Burst (FFB) scenario where cold gas accretion onto halos at $z>5$ occurs without being heated by virial shocks. In this scenario, star formation proceeds efficiently due to short free-fall times in star-forming clouds compared to feedback timescales from SNe. This high local star formation efficiency is compatible with our observed metal dilution effect because the overall ISM metallicity remains dominated by the inflowing cold gas streams, with newly produced metals not fully mixing with these streams \citep{Dekel_23,Lizhaozhou_23}. Thus, the observed lower metallicity results from enhanced metal dilution due to intense gas accretion at $z>5$. \\

In the first panel of Fig. \ref{fig:FMR_predictions}, we show the predicted FMR with model 1, where the galaxies are feed with continuous inflow with varying star formation efficiency $\epsilon_\text{SF}$. We observe that the $Z-\text{SFR}$ relation is sensitive to $\epsilon_\text{SF}$, with a lower $\epsilon_\text{SF}$ leading to a lower metallicity for a given SFR and stellar mass. This aligns with the prediction of Eq. \ref{eq:Z_obs_dilution}, which indicates that a lower $\epsilon_\text{SF}$ results in a reduced SFR, leading to a lower $\epsilon_\text{acc}$ and a stronger dilution effect. In the second panel of Fig. \ref{fig:FMR_predictions}, we show the predictions from model 2, where no inflows are available, but the galaxies can consume the gas pre-existing in the gas reservoir. We find a similar trend of FMR evolution with $\epsilon_\text{SF}$. The high rate of continuous inflows acts similarly to the pre-existing pristine gas reservoirs, as in both cases, the galaxies are in a non-equilibrium state due to abundant gas supply at the beginning of star formation. But model 2 is less realistic: the assumption of massive initial gas reservoirs without prior star formation is unlikely, as dark matter halos with masses $M_\text{h}\gtrsim10^{8}M_\odot$ already start to cool and collapse to boost star formation \citep{Klessen_23}. Nevertheless, both models highlight the effects of the delay of metal enrichment due to gas dilution in shaping the FMR evolution, and a lower $\epsilon_\text{SF}$ leads to extra metal deficiency.

\subsubsection{Feedback}
The feedback strength impacts the metal retention in galaxies. Eq. \ref{eq:Z_obs} indicates that the observed metallicity both depends on mass loading factor $\eta$ and the outflow response of star formation $\lambda$. From Eq. \ref{eq:Z_obs}, an enhanced $\lambda$ leads to a steeper slope of FMR, which leads to a lower metallicity for a given SFR and stellar mass. The mass loading factor $\eta$ sets the normalization of FMR. A higher $\eta$ leads to an enhanced metal loss and lowers the metallicity intercept of FMR. \\

In the third and fourth panels of Fig. \ref{fig:FMR_predictions}, we show the predicted FMR with model 3 and model 4, with galaxies fed by continuous inflow and varying mass loading factor $\eta$ and outflow mode $\beta$. The FMR is sensitive to both parameters. Higher $\eta$ produces lower metallicity at given SFR and stellar mass, consistent with Eq. \ref{eq:Z_obs}, where higher $\eta$ causes stronger metal loss. For model 3, metallicity plateaus at low SFRs indicate equilibrium states where metal production balances metal loss. These plateaus are higher in metallicity with lower $\eta$ due to less efficient winds. Model 4 shows similar behavior, with plateaus at low SFRs where $\beta$ has maximum impact. Varying $\beta$ changes the dependence on halo mass and modifies the outflow gas content.

In observations, however, \citet{Zhang_24,Xuy_25} suggest less efficient outflows at $z>6$ with low wind velocities. Such kind of low-velocity winds are likely producing fountain-type outflows, where the ejected gas and metals cannot escape from the gravitational potential well and eventually reaccrete onto the galaxy. The reaccretion of pre-enriched streams aids galaxies in retaining their metals \citep{Zhang_23}. On the other hand, the stellar yield $y$ also affects the normalization of FMR. An evolving initial mass function (IMF) can lead to different stellar yields \citep{vanDokkum_08}, e.g., a top-heavy IMF results in a higher total metal yield compared to a standard or bottom-heavy IMF due to increased proportion of massive stars \citep{Vincenzo_16}. From recent JWST observations, \citet{Zou_24,Hutter_25} suggest a top-heavy IMF at $z>6$ using chemical abundance patterns and UV luminosity functions. While we expect a higher metal yield with such a top-heavy IMF, those constraints on feedback cannot solely explain the metal deficiency at $z>5$ in our observations. Thus, the models and observations suggest that feedback is not the dominant process driving metal deficiency at $z>5$.

\subsubsection{Environmental effects}

\begin{figure}[t!]
    \centering
    \begin{tabular}{c}
        \includegraphics[width=8.5cm]{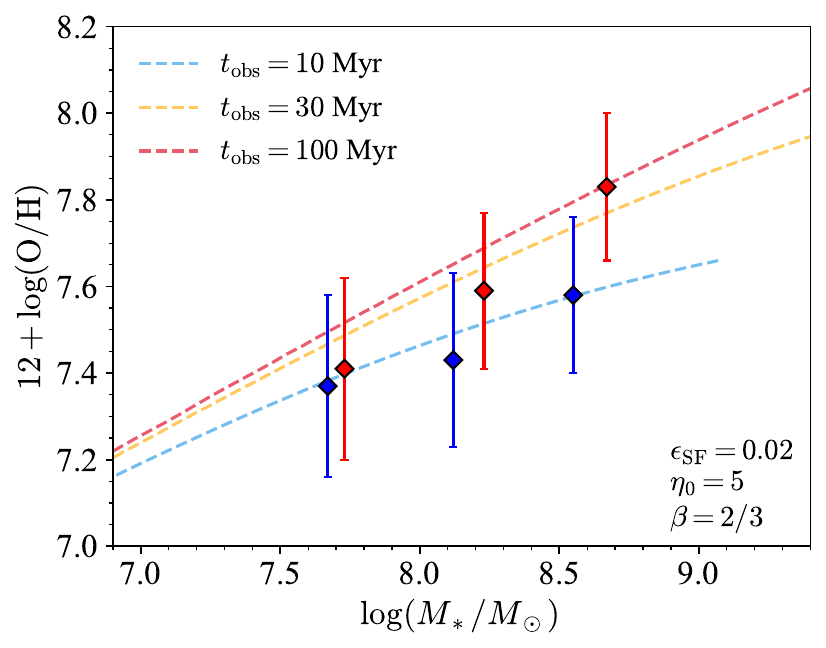} \\
        \includegraphics[width=8.5cm]{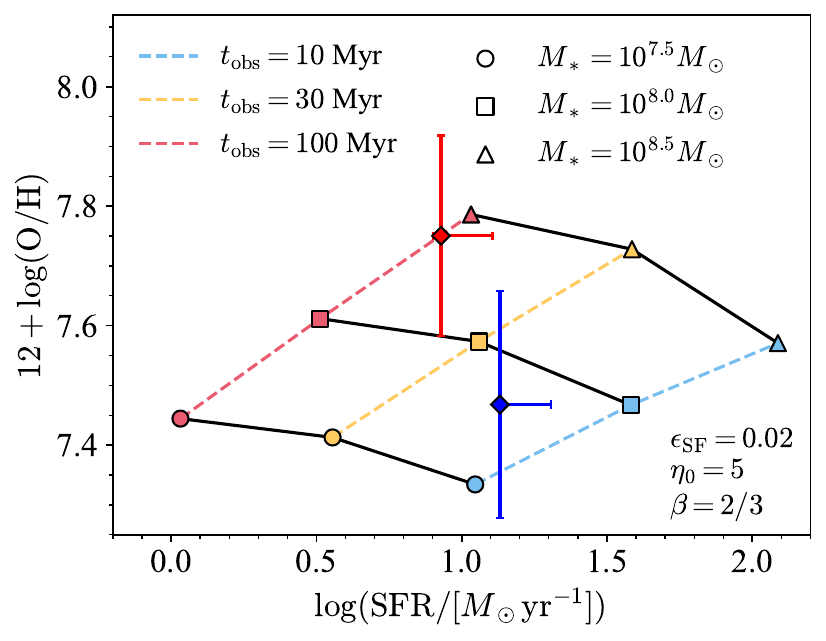}
    \end{tabular}
    \caption{Top: The predicted MZR for galaxies with different $t_\text{obs}$. The dashed lines are MZRs color-coded by $t_\text{obs}$. This model assumes $\epsilon_\text{SF}=0.02$, $\eta_0=5$, and an energy-driven wind $\beta=2/3$. The diamond points are our observations, the same as in Fig. \ref{fig:mzr}. Bottom: The predicted SFR$-$Z relation for galaxies is shown for different stellar masses and $t_\text{obs}$. The dashed lines, color-coded by $t_\text{obs}$, connect galaxies with the same observational time, while the solid black lines link galaxies of equal stellar mass. Different stellar masses are represented by different symbols. The red and blue diamond points are our observations in cluster and field environments. The asymmetric errors on SFR are due to uncertainties in dust correction. }
    \label{fig:MZR_predictions}
\end{figure}

Eq. \ref{eq:Z_obs_dilution} indicates that there is a delay of metal enrichment compared to the steady-state solution due to faster gas accretion than gas depletion. \citet{Helton_24} found that cluster galaxies at $z\sim5$ have earlier star formation, by comparing SFR at different epochs of star formation history. They found that galaxies in an overdensity show both a higher $\text{SFR}_{0-100~\text{Myr}}$ and $\text{SFR}_{30-100~\text{Myr}}/\text{SFR}_{0-30~\text{Myr}}$ ratio than their mass-matched field counterparts. They indicate that overdensity galaxies have already undergone significant star formation at $30-100$ Myr before observation, while the major star formation for field galaxies just started in the last 30 Myr. Estimated from the SED fitting, we find that the stellar masses observed in the overdensities are more massive by $\sim0.1$ dex than field galaxies. 
In contrast, their median \OIII\ and \Hb\ luminosities are $\sim0.1$ dex lower than those of the field. Since \Hb\ emission traces the instantaneous SFR, this suggests that field galaxies are forming stars more actively, consistent with the scenario of enhanced recent star formation in the field. In addition, weaker nebular emissions like \OIII\ in overdensities also indicate more evolved systems. However, because the \OIII\ and \Hb\ lines are sensitive to dust, and \citet{Liq_25} suggested that galaxies are more dusty in the overdensities, we caution possible uncertainties in \OIII\ and \Hb\ fluxes and derived SFRs for the two samples.

Supposing the cluster galaxies are more evolved, they should be closer to equilibrium with a higher observed metallicity close to the steady-state solution in our models.
We further investigate such effects from model 5. Using the model output grids, we select galaxies at different times since formation $t_\text{obs}=[10,30,100]$ Myr, which can alternatively be interpreted as the ``age'' of the galaxy.
In the top panel of Fig. \ref{fig:MZR_predictions}, we show the predicted MZR for galaxies with different $t_\text{obs}$. We find that galaxies with longer durations of star formation have higher metallicities. The galaxies observed at $t_\text{obs}=10$ Myr show a flatter MZR slope, while those observed at $t_\text{obs}=100$ Myr show a steeper slope with $\sim0.2$ dex higher metallicities. The steeper slope can be predicted by Eq \ref{eq:Z_steady}. We have adopted $\lambda=1$ in the model, and now we approximate $\eta\gg1$. In a steady-state solution, the slope of MZR can be estimated by:
\begin{equation}
    d\log(Z_\text{ISM})/dM_* = - d\log(\eta)/dM_*=\beta\cdot\log(\mathcal{F})',\label{eq:slope}
\end{equation}
where $\log(\mathcal{F})'$ is the derivative of Halo-to-Stellar mass function in log space. For halos at $z\sim5$, masses lower than $M_h=10^{12}M_\odot$, $\log(\mathcal{F})'$ can be approximated as a constant $\approx0.6$ \citep{Shuntov_22}. As a result, the MZR slope\footnote{\citet{Guo_16} estimate $\log(\mathcal{F})'\approx0.5$ from \citet{Moster_10,Behroozi_13}, and they predict slightly different slopes with the same equation.} is estimated as $\approx0.6~\beta$ for low mass galaxies at $z\sim5$. For energy-driven wind used in our model, we expect a slope of $\sim0.4$ when reaching steady-state conditions. For galaxies with longer time to evolve, i.e., the solutions with $t_\text{obs}=100$ Myr, they are closer to equilibrium, and the observed MZR slope is close to the steady-state approximation.
This indicates that the MZR can vary with time during non-equilibrium phases for galaxies with ongoing gas accretion. In addition, our model suggests energy-driven wind with mechanical feedback to produce the observed MZR in the non-equilibrium phase. A momentum-driven wind with $\beta=1/3$ is also possible, but it requires the steady-state condition for all the galaxies to match the predicted slope $\gamma\approx0.6\times1/3=0.2$.\\

In the bottom panel of Fig. \ref{fig:MZR_predictions}, we show the predictions of the Z-SFR relation from the same output of model 5. We see that at the same time of observation, more massive galaxies have higher metallicities, which is expected by MZR predictions. We also find that FMR evolves with time; the contours denoting the observation time are moving upwards as time increases. This indicates that the FMR evolves with time, and the different stages of star formation history can also be a driver of FMR. We also compare our observations in the bottom panel of Fig. \ref{fig:MZR_predictions}. We find that the $Z-\text{SFR}$ relations for cluster and field samples are not moving along the predicted contour of stellar mass at fixed observation time, while they cross the contours of both stellar masses and observation time. 
The model predictions are consistent with the case that cluster galaxies are more massive and older than field galaxies.
% consistent with results from SED fitting \citep{Champagne_25}. 
Thus, the model suggests a later evolution phase for cluster galaxies and a younger stage for field galaxies.
If cluster galaxies have earlier star formation, we expect them to be more metal-rich than field galaxies on the FMR diagram from our model predictions. Our observation thus suggests that cluster galaxies have an earlier onset of star formation within a non-equilibrium scenario, consistent with the finding of \citet{Helton_24}.

\section{Discussion}\label{sec:discussion}

\subsection{Comparison with hydrodynamical simulations}

We recognize that our analytical model is oversimplified. Any small-scale processes are not modeled, such as gas dynamics including rotation, turbulence, diffusion, and advection \citep[e.g.,][]{Sharda_21}. The ISM gas-phase structures are not considered, and the more detailed processes in feedback, such as heating and cooling, are not included either \citep[see, e.g.,][]{Collacchioni_18}. And the star formation efficiency can be modulated by feedback, which is not considered in our model.
We have also assumed that the inflowing gas is metal-free $Z_{\text{in}}=0$. However, this may not always hold true in realistic environments. The IGM can be enriched by earlier generations of galaxies, and inflowing gas may also contain recycled material from the circumgalactic material (CGM) that has been ejected and subsequently reaccreted \citep{Beckett_24}, especially if the outflows are not powerful enough to escape the galaxy's gravitational potential. Simulations and observations suggest that the CGM of high-redshift galaxies is often enriched, and the metallicity of inflows may depend on cosmic time, halo mass, and the surrounding galaxy environment \citep{Angles_17}. As such, the inflow metallicity may vary with environment, star formation rate, and dynamical timescales, and could potentially correlate with \( Z_{\text{out}} \) and \( \dot{M}_{\text{out}} \), though this relation is non-trivial in practice.

\begin{figure}[t!]
    \centering
        \includegraphics[width=8.5cm]{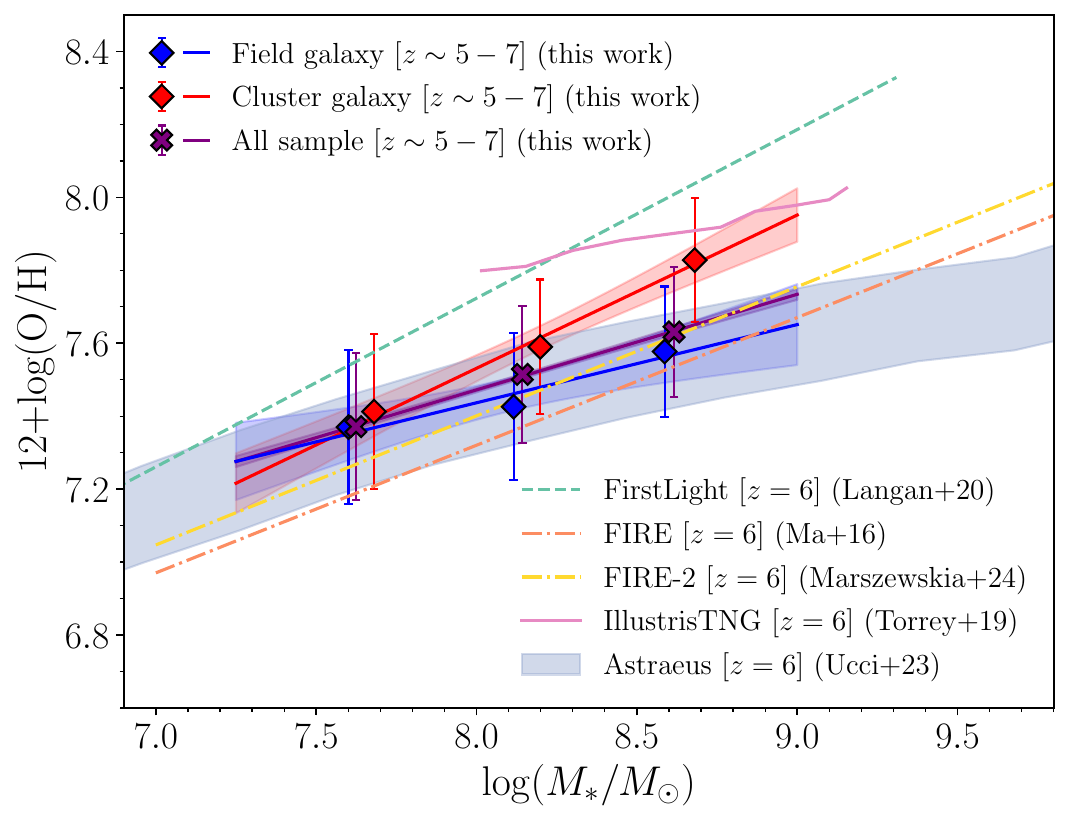} \\
    \caption{The observed MZR compared with hydrodynamical simulations. The symbols for our observations are the same as in Fig. \ref{fig:mzr}. The results at similar redshifts from different suites of hydrodynamical simulations are denoted by different colors, including \textsc{FirstLight} \citep{Langan_20}, FIRE \citep{Ma_16}, FIRE-2 \citep{Marszewski_24}, \textsc{IllustrisTNG} \citep{Torrey_19}, and \textsc{Astraeus} \citep{Ucci_23}.}
    \label{fig:MZR_sim}
\end{figure}

Thus, comparing observations with numerical simulations helps understand the physical processes. 
Various hydrodynamical simulations have studied metal enrichment in galaxies, including \textsc{EAGLE} \citep{Rossi_17}, \textsc{FIRE} \citep{Ma_16}, \textsc{FIREbox} \citep{Bassini_24}, \textsc{FIRE-2} \citep{Marszewski_24}, \textsc{IllustrisTNG} \citep{Torrey_19}, \textsc{FirstLight} \citep{Langan_20}, \textsc{Astraeus} \citep{Ucci_23}, \textsc{SERRA} \citep{Pallottini_24}, and \textsc{FLARES} \citep{Lovell_21}. In Fig. \ref{fig:MZR_sim}, we compare our observations with these simulations. Our full sample's MZR is generally consistent with \textsc{FIRE}, \textsc{FIRE-2}, and \textsc{Astraeus} simulations, while \textsc{FirstLight} and \textsc{IllustrisTNG} predict higher metallicity normalizations.
However, metallicity definitions vary between simulations. For example, \citet{Marszewski_24} defines gas-phase metallicity as the mass-weighted mean metallicity of gas particles within 0.2 virial radius, while \citet{Ma_16} used 0.1 virial radius and required gas temperatures below $10^4$ K. Different metal yield prescriptions also affect the absolute metals produced in simulations. For instance, \citet{Langan_20} used yields from \citet{Woosley_95} and defined solar metallicity as $12+\log(\text{O/H})=8.9$, higher than in other studies like \citet{Torrey_19}.
Therefore, the MZR slope is more informative than its absolute normalization, as it traces relative metal enrichment across stellar masses, reflecting the competition between star formation, inflows, and outflows. Fig. \ref{fig:MZR_sim} shows that \textsc{Astraeus} and \textsc{IllustrisTNG} simulations best match our observed slopes. \textsc{FIRE} and \textsc{FIRE-2} predict steeper slopes but remain within $1\sigma$ errors, while \textsc{FirstLight} predicts a steeper slope closer to our observations in overdense environments.

\citet{Ibrahim_24} investigated the impact of different feedback mechanisms on MZR using hydrodynamical simulations based on \textsc{gadget-3} code. They find that mechanical feedback best reproduces observations up to redshift $z\sim3$. Our analytical model also suggests the presence of mechanical feedback at $z\approx5$–$7$, although it alone is insufficient to reproduce the observations. From \textsc{FIREbox} simulation, \citet{Bassini_24} suggest that the evolution of MZR at $z\lesssim3.5$ is driven by inflow metallicity, outflow metallicity, and mass loading factor, rather than gas fraction \citep[e.g.,][]{Rossi_17,Torrey_19}. On the other side, \citet{Marszewski_24} found a weak evolution of MZR at $z\gtrsim5$ in FIRE-2 simulations with a non-evolving slope $\gamma=0.37$. They find that the gas fractions are mass-dependent and vary substantially with redshift, which indicates gas fractions alone cannot explain weakly evolving MZR, and inflows and outflows must be taken into account. \citet{Angles_17} suggest that recycled gas also plays a crucial role in shaping galaxy metallicities across different stellar masses, with low-mass galaxies being more likely to accrete pre-enriched gas, while high-mass galaxies primarily accrete pristine gas. \citet{Langan_20,Ma_16} found that a ``leaky box" model can reproduce the MZR evolution at $z\gtrsim3.5$ in simulations but it requires an effective yield ($y_\text{eff}$) an order of magnitude smaller than intrinsic stellar yield $y$, highlighting the non-negligible impact of inflows and outflows. 

\citet{Wangk_23} have investigated the environmental effect on MZR at $z\lesssim2$ using EAGLE simulation. They found that both the suppression of gas accretion and the ram pressure stripping contribute to the environmental dependence of satellite galaxies at $z\sim2$. The ram pressure strips the ISM gas as satellite galaxies move across the clusters and come into contact with the hot ICM. It leads to starvation of the satellite galaxies, preventing further gas accretion, and the galaxies become more metal-rich.
Additionally, they observed that central galaxies in protoclusters at $z\sim2$ tend to reside in more massive halos, accrete more gas through cold-mode accretion, and consequently exhibit lower metallicities. 
% More detailed study on protocluster properties at $z>5$ are required. 

On the other hand, directly comparing our observations in overdense environments with simulations at $z>5$ remains highly challenging. Simulating low mass galaxies at high redshifts requires high resolution of particles (e.g., baryon particle mass $<10^5 M_\odot$), while simulating the large scale structures requires a large simulation box (e.g.,$\sim$ Gpc scale). At present, hydrodynamical simulations capable of meeting the requirements of both large and small scales have yet to be fully developed. For example, recent \textsc{TNG-Cluster} simulations have provided simulated galaxies in massive clusters within a $\sim1$ Gpc$^3$ volume \citep{Nelson_24}, but their baryon particle of $\sim10^7 M_\odot$ is too massive to resolve low mass galaxies.

\subsection{Implication of environmental effect on MZR}\label{sec:implication}
Our analytical models demonstrate that the differences in MZR and FMR between field and cluster galaxies can be explained by delayed metal enrichment, primarily caused by the dilution effect of gas accretion. In overdense environments, earlier star formation and rapid metal enrichment may be enhanced compared to blank fields.

Environmental effects on galaxy metallicities remain debated in the literature. For example, \citet{Wang_22,Li_22} found a flatter MZR in a massive protocluster at $z = 2.24$, indicating that massive protocluster galaxies are more metal-poor than their field counterparts. Similarly, \citet{Calabro_22} observed more metal-poor galaxies located in denser environments compared to those of similar masses in underdense regions. In support of these findings, EAGLE simulations \citep{Wangk_23} suggest that low-metallicity gas accretion dilutes the metallicities of massive central galaxies, while gas stripping suppresses cold gas accretion in lower-mass satellites—consistent with the observations in \citet{Wang_22}.

Ram pressure stripping is a possible, but uncertain, mechanism at $z > 5$. At $z \sim 2$, metal enhancement in protocluster galaxies has been observed by \citep{Perez-Martinez_23,Shimakawa_15}, who attribute it to more efficient gas recycling or gas stripping \citep{Kulas_13}. While gas stripping could contribute to metal enhancement in protocluster galaxies, the early stage of protocluster formation implies that their halos are not yet virialized, and the ICM is expected to remain relatively cool. Whether the ICM is sufficiently hot and dense to support ram pressure stripping and suppress gas accretion remains unclear \citep{Jachym_07,Boselli_14,van_de_Voort_17}. The evidence for ram pressure stripping at the highest redshifts is currently up to $z = 2.51$ \citep{Xu_25}, leaving its effectiveness at $z > 5$ largely unexplored.

Strangulation might still play a role in $z > 5$ protoclusters. The gravitational potential of the protocluster halo could induce tidal stripping of ISM gas, inhibiting further gas inflow and leading to enhanced metallicity \citep{Peng_15}. This scenario is consistent with our observations: in our non-equilibrium framework, galaxies that evolve closer to chemical equilibrium exhibit higher metallicities, reflecting a later evolutionary phase. If gas accretion is suppressed via tidal effects in overdense environments, equilibrium between metal production and dilution can be achieved earlier. However, we note that while this explanation aligns with our observations, direct observational evidence for strangulation remains lacking.

Whether gas accretion or gas stripping dominates in overdense environments may also be mass-dependent. Studies at $z \sim 2$ suggest that massive galaxies are more affected by gas accretion, while low-mass galaxies are more sensitive to gas stripping \citep{Wangk_23}, although the stellar mass threshold between the two mechanisms remains uncertain. As shown in Fig. \ref{fig:d_oh}, metallicity deviations between overdense and field environments across different studies are different, even when their galaxy samples fall within the same mass and redshift range. One explanation is that environmental effects depend on the nature of the overdensity itself; redshift, halo mass, and virialization state may all relate to the environmental effects. For example, early-stage protoclusters at high redshift may have different effects than virialized clusters at lower redshift: the effect of cold gas accretion may be active in early protoclusters, while the gas supply could be suppressed by the hot halo of virialized clusters.

\citet{Chiang_17} outlined three major phases of cluster evolution. In the early phase ($z \gtrsim 5$), protoclusters begin forming in the densest regions through an "inside-out" process, with accelerated central star formation. The intermediate phase ($z \sim 5-1.5$) involves rapid stellar mass assembly and structural growth, while the late phase ($z < 1.5$) is marked by satellite accretion and gradual suppression of star formation. Our protocluster candidates at $z > 5$ are likely in the early phase, undergoing active mass assembly with significant variation in their properties.
As noted by \citet{Perez-Martinez_23}, even protoclusters at the same redshift may be at different evolutionary stages, leading to diverse environmental effects. Thus, combining all overdensities into a single subsample may average meaningful variations \citep{Morishita_25}. Case-by-case investigations of individual protoclusters at $z>5$ are subject to future study \citep{Li_25c}.
% (Li et al. In prep)
In short, the environmental effects on the MZR are complex. The enhanced metallicities we observe in $z > 5$ protoclusters are consistent with accelerated star formation in overdense regions \citep{Helton_24,Morishita_25c} and align with our non-equilibrium chemical evolution models. Strangulation due to tidal effects is also a plausible contributor to faster metal enrichment \citep{Peng_15}. Ram pressure stripping could enhance metallicities, but its role remains uncertain given the likely cooler ICM conditions in early protoclusters at $z > 5$.

\subsection{Caveats in MZR measurements}\label{sec:caveats}

Given the uncertain systematics associated with different metallicity measurements (e.g., strong line calibrations and $T_e$ method) and different line indicators (e.g., R3, Ne3O2, N2\Ha) used across various studies, the origin of the discrepancy in MZR slopes remains elusive. In this study, we only used the R3 line ratio to derive the metallicity and assumed the low-metallicity solution of metallicity. There are potentially higher metallicity galaxies calculated as low metallicity ones. In addition, we are using stacked spectra to derive the MZR. The intrinsic scatter of the MZR relation cannot be estimated from the median measurements. In addition, we are only estimating the median from the stacked samples, so that the actual range of metallicities is still not constrained. While the stacked measurements are representative of the median properties, and the observed trends from stacks appear real, uncertainties in both the metallicity calibrations and the measurements imply that the differences remain compatible within the errors. Calibrations such as R3 could be affected if galaxies with extreme star formation episodes dominate the stack, especially since our sample is selected by \OIII emission and may be biased toward stronger emission-line galaxies compared to the NIRSpec sample used for calibrating the R3 relation.

Furthermore, the measurements of stellar masses also introduce systematic uncertainties in MZR. \citet{Whitler_23} shows that the assumption of SFH significantly influences the determination of stellar masses during SED fitting. 
Since young stars dominate the observed rest-UV and optical SED, they can outshine any older stellar populations present in the galaxy. As a result, the choice of the SFH model significantly impacts the derived properties, particularly the early star formation history, stellar age, and total stellar mass of the system.
Different literature studies have different assumptions of SFH, and this can introduce systematic uncertainties in stellar mass measurements and thus the derived MZR. \citet{Nakajima_23} and \citet{Heintz_23} assumed a non-parametric SFH. \citet{Chakraborty_24} and \citet{Sarkar_25} assumed a constant SFH. \citet{Curti_24} assumed a delayed-exponential SFH. 

However, as \citet{Curti_24} incorporates samples from \citet{Nakajima_23}, they mix galaxies analyzed with different SFH assumptions. \citet{Heintz_23} notes that non-parametric SFH typically increases stellar mass estimates by $0.5-1.0$ dex compared to parametric approaches (constant or delayed-exponential SFH). This mixing of "high mass" galaxies from non-parametric SFH with "low mass" galaxies from parametric SFH introduces systematic uncertainties in the MZR. For a sample with similar metallicities, higher stellar mass estimates will shift points rightward on the MZR plot, potentially flattening the slope and possibly explaining the flatter MZR in \citet{Curti_24}. While using consistent SFH assumptions reduces impact on the slope, it still affects normalization (e.g., a $\sim0.5$ dex increase in stellar masses reduces the MZR intercept by $\sim0.1$ dex at fixed slope $\gamma=0.2$). Comparisons between different MZR studies should therefore be interpreted cautiously.

There are also other sources of uncertainty in the measurements of stellar masses. In our sample, we only have three photometric bands in F115W, F200W, and F356W. The SED shape cannot be fully constrained, especially the lack of photometric coverage redward of $4000\text{\AA}$ break.
\citet{Liq_24} showed that stellar mass and SFR uncertainties can increase by $\sim 0.1$ dex without such constraints. Thus, this introduces systematic uncertainties in the derived MZR.

\subsection{Future prospects}

Due to current data limitations, such as the incomplete spectral coverage of the F356W grism, the impacts of environmental effects remain insufficiently explored. Distinguishing between these scenarios would benefit from diagnostics based on additional emission lines. 

For example, unlike estimating O/H abundances, observational estimates of N/O can be obtained without introducing significant bias from the ionization structure \citep{Perez-Montero_09,Peng_21,Perez-Diaz_25}. Furthermore, the metallicity estimated from $12+\log(\mathrm{O/H})$ can be influenced by hydrodynamical processes that change the total gas content. For example, stochastic gas inflows in evolved systems may temporarily dilute their metallicity, causing them to appear as unevolved systems, while the $\log(\mathrm{N/O})$ should reflect the stellar chemical enrichment purely. Thus, the accelerated star formation scenario should also be expected from the $\log(\mathrm{N/O})$ ratio, pointing towards higher values than field galaxies, if they had enough time for the secondary enrichment through CNO cycles. From a small sample, \citet{Morishita_25c} tentatively found that galaxies in two overdensities at $z\sim5.7$ exhibit a higher fraction of weak H$\alpha$ equivalent widths (EW) and strong 4000 \AA\ continuum breaks, which are characteristic features of evolved stellar populations. A more robust test of the accelerated star formation scenario requires a larger sample with measurements of H$\alpha$ equivalent widths and N/O ratios to probe the stellar chemical enrichment.

The alternative scenario of gas stripping could be distinguished through interstellar medium conditions probed by different emission-line diagnostics. \citet{Zhou_25} found enhanced \OI/\Ha ratios in protocluster member galaxies at $z \sim 2$, suggestive of increased shock excitation possibly driven by gas stripping. Similar tests could be applied to earlier galaxies to investigate whether such processes are in effect at $z \sim 6$.

\section{Summary} \label{sec:summary}
In this study, we have investigated the MZR and FMR using a parent sample of 604 galaxies spanning a stellar mass range $10^7<M_*/M_\odot<10^9$ and at redshift $5<z<7$, observed with JWST NIRCam WFSS. The sample comprises \OIII emitter identified in the 25 quasar fields in ASPIRE survey and one quasar field, SDSS J0100+2802, from the EIGER survey. We have identified significant clustering of galaxies around quasar redshifts, as well as additional galaxy clustering at random redshifts. We have divided the sample into field and overdense environments and measured the MZR and FMR using stacked spectra. Our analysis reveals distinct metallicity properties between field and cluster galaxies. We explore the underlying physical processes producing such a difference with an analytical model.

Our main findings are summarized as follows:
\begin{enumerate}
    \setlength{\itemsep}{0.5em}
    \item We estimate the electron temperature of $2.0^{+0.3}_{-0.4}\times10^4$ K from the \Hg\ and $\OIII_{4363}$ in the stacked spectrum, indicating a metal-poor sample with median gas phase metallicity 12+$\log(\mathrm{O/H})=7.65^{+0.26}_{-0.15}$. Dust attenuation derived from the Balmer decrement is $A_V=0.38^{+0.54}_{-0.38}$, suggesting a moderate dust attenuation in star-forming regions.
    \item We explore the MZR of our full sample at $z=5-7$. The MZR slope is $\gamma = 0.26 \pm 0.01$, consistent with recent studies at similar redshifts \citep[e.g.,][]{Chakraborty_24,Sarkar_25}.
    \item Galaxies in overdense environments exhibit a steeper MZR slope ($\gamma = 0.42 \pm 0.04$) compared to field galaxies ($\gamma = 0.21 \pm 0.06$), and they are more metal-rich, especially at higher masses $\log(M_*/M_\odot) > 8$. We build a mass-matched sample of field galaxies and still find that the galaxies in overdense environments are more metal-rich by $\sim0.2$ dex than their field counterparts with the same stellar masses.
    \item We compare the FMR of our sample galaxies with the local FMR using $\alpha=0.66$ \citep{Andrews_13}. We find that our sample galaxies are metal-deficient by $0.2$ dex compared to the local FMR. However, the galaxies in overdense environments align closer to the local FMR predictions.
    \item  We apply a bathtub model to explain the MZR/FMR differences between field and cluster galaxies. Our model shows that galaxies actively accreting gas exist in a non-equilibrium state with lower metallicity than equilibrium predictions. The time evolution in our model qualitatively explains our observations, suggesting that gas accretion dilution dominates metallicity evolution at $z>5$. More evolved systems approach equilibrium and experience less dilution. This framework explains the enhanced metallicity in overdense environments as resulting from earlier star formation, giving these galaxies more time to approach chemical equilibrium.
    \item We also compare the MZR of our sample with the results from hydrodynamical simulations. We find that the observed MZR for our full sample is generally consistent with the results from FIRE \citep{Ma_16}, FIRE-2 \citep{Marszewski_24}, and \textsc{Astraeus} \citep{Ucci_23} simulations. Those simulations also emphasize the influence of gas accretion and feedback in shaping the metallicity evolution at $z>5$.
    \item We additionally compare the environmental effects with literature observations and simulations. We find that the enhanced metallicity in overdense environments can be related to accelerated star formation in protoclusters at $z>5$ \citep{Helton_24}. Strangulation in overdense environments is also plausible. Ram pressure stripping can also lead to metal enhancement in protoclusters \citep{Wangk_23}, but it is not clear if the less virialized protoclusters at $z>5$ can sustain hot ICM to support ram pressure stripping.
\end{enumerate}

Our results emphasize the critical role of the environment in regulating early chemical enrichment. Future JWST/NIRSpec follow-up will refine metallicity diagnostics/measurements and explore variations in ionization conditions, while deeper and wider grism surveys will extend these studies to lower masses and larger areas. For example, the ongoing JWST COSMOS-3D survey will provide a unique opportunity to study the metal-enrichment in a much larger area, mapping both underdense and overdense structures in the COSMOS field \citep[e.g.,][]{Horowitz_22}.

\begin{acknowledgements}
        We are grateful to the anonymous referee for their constructive feedback, which significantly strengthened the paper. Z.L. thanks Kai Wang, Zhaoran Liu, and Kasper Heintz for helpful discussions, and Kimihiko Nakajima for help with SED parameters. Z.L. acknowledges the fellowship from the Cosmic Dawn Center. The Cosmic Dawn Center is funded by the Danish National Research Foundation under grant no. 140. K.K. acknowledges support from VILLUM FONDEN (71574). L.C. is supported by DFF/Independent Research Fund Denmark, grant-ID 2032–00071. E.P.F. is supported by the international Gemini Observatory, a program of NSF NOIRLab, which is managed by the Association of Universities for Research in Astronomy (AURA) under a cooperative agreement with the U.S. National Science Foundation, on behalf of the Gemini partnership of Argentina, Brazil, Canada, Chile, the Republic of Korea, and the United States of America. We acknowledge the use of Astropy \citep{Astropy}, \textsc{Grizli} \citep{Brammer_2022}, \texttt{Calwebb} \citep{bushouse_2024_10870758}, Scipy \citep{scipy}, \texttt{BEAGLE} \citep{Chevallard_16}, \texttt{SourceXtractor++} \citep{sepp}, and \textsc{Matplotlib} \citep{Matplotlib}. This work is based on observations made with the NASA/ESA/CSA James Webb Space Telescope. The JWST data presented in this article were obtained from the Mikulski Archive for Space Telescopes (\href{http://archive.stsci.edu}{MAST}) at the Space Telescope Science Institute, which is operated by the Association of Universities for Research in Astronomy, Inc., under NASA contract NAS 5-03127 for JWST. These observations are associated with JWST programs \#2078 and \#1243.
\end{acknowledgements}

\section*{Data Availability}
The JWST data are available in the Mikulski Archive for Space Telescopes (MAST; \url{http://archive.stsci.edu}), under JWST programs GO-2078 (\url{https://doi.org/10.17909/vt74-kd84}) and GO-1243 (\url{https://doi.org/10.17909/m5mp-5v90}).

\bibliographystyle{aa}
\bibliography{sample631}

\begin{appendix}

\section{Stacking bias}\label{sec:bias}
\begin{figure}[t!]
    \centering
    \resizebox{8.5cm}{!}{\includegraphics{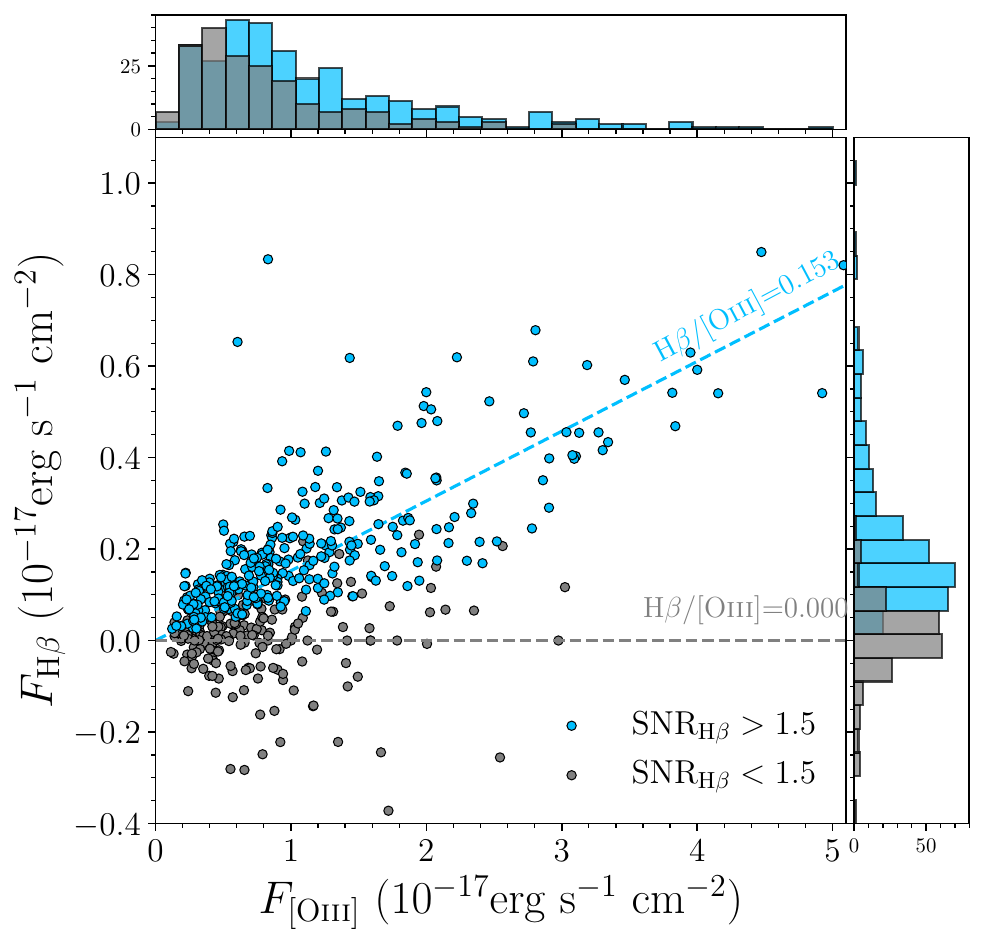}}
    \caption{The flux distribution of \OIII and \Hb\ in our sample. The blue circles represent measurements with $\text{SNR}_{\Hb}>1.5$, while the gray circles represent low-quality measurements with $\text{SNR}_{\Hb}<1.5$. The blue dashed line marks the linear regression with the $\OIII-\Hb$ relation, with the slope indicating the inverse of the median R3 ratio. Similar regression is applied to $\text{SNR}_{\Hb}<1.5$ measurements, shown in the gray dashed line. No meaningful relations on R3 are constrained by those low-quality measurements.
    % No measurable R3 ratio is detected for those low-quality measurements.
    % TBD. Draw flux distributions... are they the same? high o3 flux outliers?
    }
    \label{fig:snr}
\end{figure}

The measured $F_{\Hb}-F_{\OIII}$ relation is shown in Fig. \ref{fig:snr}. We show the data points with different colors for $\text{SNR}_{\Hb}>1.5$ and $\text{SNR}_{\Hb}<1.5$. For low $\text{SNR}_{\Hb}<1.5$ targets, the measured \Hb\ fluxes can be fitted with a normal distribution with a mean of $\sim0.01\times~\mathrm{10^{-17}~erg~s^{-1}~cm^{-2}}$. Thus, for this part of the sample, the \Hb\ signals are predominantly buried in random noise. It is difficult to measure the true \Hb\ fluxes even after stacking, as the measured positive and negative fluxes cancel out with a median around zero. Incorporating this part of the sample leads to an underestimation of \Hb\ fluxes, and the \OIII/\Hb\ ratios result in an excessively high $\OIII/\Hb\gtrsim10$. Similarly, \citet{Topping_20} have found that the best-fitting stellar parameters, including metallicity, are significantly biased if including low SNR spectra. The setting of an SNR cut before stacking is also applied in similar studies \citep{Sanders_21,Andrews_13}.

\begin{figure}[t!]
    \centering
    \resizebox{8cm}{!}{\includegraphics{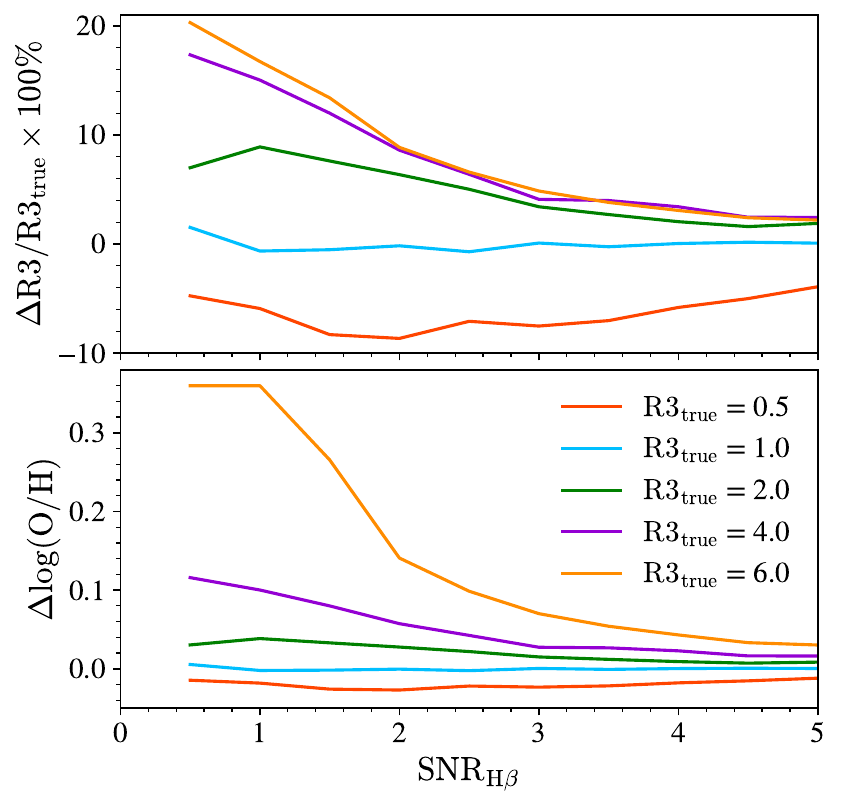}}
    \caption{The bias of R3 ratios and metallicities from median stacking for simulated spectra with different $\rm SNR_{\Hb}$. The different colors represent results with different input line ratios.}
    \label{fig:stacking_bias}
\end{figure}

Here, we further assess the impact of including low-SNR spectra in the stacking analysis. Assuming a uniform and properly subtracted background, the background noise should follow a Gaussian distribution with a zero mean $B\sim\mathcal{N}(0,\sigma^2)$, and the observed line fluxes follow Gaussian distributions centered at their true values $F\sim\mathcal{N}(F_{\text{true}},\sigma_F^2)$. 
However, when taking ratios of noisy measurements like \OIII and \Hb, the measured ratios follow a normal ratio distribution and are generally not symmetric around the true ratio values \citep{JSSv016i04}. The normal ratio distribution does not have a general closed-form solution for its mean and standard deviation.
To quantify potential biases in the stacking analysis of emission lines, we performed Monte Carlo simulations of spectral stacking with controlled input parameters, designed to replicate the observational characteristics of our dataset while maintaining known line ratios for validation.
In each $i^\text{th}$ realization, the steps are as follows. 
\begin{enumerate}
    \item Randomly resample \OIII line fluxes from our observed catalog to generate a mock dataset, preserving the actual flux distribution.
    \item Generate corresponding \Hb\ fluxes ($f_{\Hb}$) by applying a known ``true" \OIII/\Hb\ ratio $(\rm R3_{true})$.
    \item With an input $\rm SNR_{\Hb}$, the \Hb\ flux uncertainty is defined as $\sigma_{\Hb} = f_{\Hb}/\rm {SNR_{\Hb}}$. Assuming the same background noise, the \OIII flux uncertainty is set the same as $\sigma_{\Hb}$. We add Gaussian noise to both $f_{\Hb}$ and $f_{\OIII}$ measurements.
    \item Compute the observed median line ratio in this dataset $(\mathrm{R3}_{\mathrm{obs},i})$ using median flux for observed \OIII and \Hb\ lines.
\end{enumerate}
This process was repeated 1000 times for each set of input parameters to build statistical distributions of the recovered line ratios. We tested true line ratios from R3 = 3.0 to 6.0 and SNR of H-beta between 0.5 and 5.0, covering the range relevant to our observations.
The bias in the recovered line ratios is quantified as the relative error: $\rm (R3_{obs}-R3_{true})/R3_{true}$, where $\rm R3_{obs}$ is the median of the stacked ratios in 1000 random realizations. We further translated these biases into metallicity offsets using R3 calibration \citepalias{Sanders_24} to assess the impact on derived oxygen abundances. The results are shown in Fig. \ref{fig:stacking_bias}. We find a notable bias at $\rm SNR_{\Hb}<1.5$ for R3 ratios larger than two, where the observed ratio exceeds the true ratio by $\sim 10\%-20\%$, leading to a metallicity bias toward the high end by $\sim0.1-0.3$ dex. When the SNR increases, the observed line ratios converge to the true ratios. 
We also show the cases for small ratios with $\rm R3_{true}=0.5,1$. They have zero bias only when they are equal, and a negative bias for ratios smaller than one. The low ratio cases may not reflect the realistic conditions for our sample, as they often require too low metallicities, e.g., $\rm R3<1$ for $12+\log(\rm O/H)<6.5$, assuming \citetalias{Sanders_24} calibrations. They can also reflect the cases for other emission line ratios such as $\Hg/\Hb=0.47$. 

Thus, we caution the potential bias of measuring line ratios for low SNR spectra even after median stacking. Such bias can be mitigated with higher SNR spectra.

\section{Summary of sample properties}\label{sec:table}

\begin{table*}[t]
\renewcommand{\arraystretch}{1.35}
\caption{The R3 diagnostics and inferred metallicity of in stacks of different subsets of our sample (with \Hb\ coverage at $z>5.5$) split by stellar mass.}
\label{tab:r3_info}
\centering
\begin{tabular}{cccccc}
\hline
\hline
 & $\rm \log(M_*/M_\odot)$ & R3 & $^{a}$12+$\rm \log(O/H)_{C24}$ & $^{b}$12+$\rm \log(O/H)_{S24}$ & $^{c}$$N_\text{gal}$ \\
\hline
\multirow{3}{*}{All (Mass bin)} & $7.68^{+0.13}_{-0.24}$ & $4.61\pm0.47$ & $7.37\pm0.20$ & $7.31\pm0.21$ & 74\\
 & $8.16^{+0.13}_{-0.17}$ & $5.42\pm0.40$ & $7.51\pm0.19$ & $7.45\pm0.20$ & 91\\
 & $8.57^{+0.27}_{-0.15}$ & $5.94\pm0.47$ & $7.63\pm0.18$ & $7.56\pm0.18$ & 95\\
\hline
\multirow{3}{*}{Cluster (Mass bin)} & $7.73^{+0.13}_{-0.23}$ & $4.81\pm0.74$ & $7.41\pm0.21$ & $7.35\pm0.21$ &28 \\
 & $8.23^{+0.10}_{-0.19}$ & $5.73\pm0.73$ & $7.59\pm0.18$ & $7.51\pm0.19$ & 33\\
 & $8.67^{+0.18}_{-0.20}$ & $6.41\pm1.03$ & $7.83\pm0.17$ & $7.69\pm0.17$ & 24\\
\hline
\multirow{3}{*}{Field (Mass bin)} & $7.67^{+0.13}_{-0.24}$ & $4.57\pm0.65$ & $7.37\pm0.21$ & $7.30\pm0.22$ & 47\\
 & $8.12^{+0.14}_{-0.16}$ & $4.95\pm0.43$ & $7.43\pm0.20$ & $7.36\pm0.20$ & 57\\
 & $8.55^{+0.27}_{-0.14}$ & $5.69\pm0.53$ & $7.58\pm0.18$ & $7.50\pm0.19$ & 64\\
\hline
Full Cluster & $8.23^{+0.46}_{-0.51}$ & $6.29\pm0.55$ & $7.75\pm0.16$ & $7.64\pm0.17$ & 85\\
\hline
Mass-Matched Field & $8.22^{+0.48}_{-0.50}$ & $5.37\pm0.34$ & $7.51\pm0.19$ & $7.44\pm0.19$ & 85\\
\hline
\end{tabular}
\tablefoot{$^{a}$ Metallicity measured with \citetalias{Chakraborty_24} calibration. $^{b}$ Metallicity measured with \citetalias{Sanders_24} calibration. $^c$ Number of galaxies in each bin.}
\end{table*}

\begin{table*}[t]
\renewcommand{\arraystretch}{1.2}
\caption{Comparing MZR with different studies in literature. For measurements in this work, we show both MZR measurements based on \citetalias{Sanders_24} and \citetalias{Chakraborty_24} calibrations.}
\label{tab:mzr_info}
\centering
\begin{tabular}{cccccc}
\hline\hline
Study & Redshift & $\log(M_*/M_\odot)$ & $\gamma$ & $Z_{10}$ & Calibration \\
\hline
This work (all) & $5-7$ & $7.0-9.0$ & $0.26\pm0.01$ & $8.00\pm0.01$ & \multirow{3}{*}{R3 \citepalias{Chakraborty_24}} \\
This work (cluster) & $5-7$ & $7.0-9.0$ & $0.42\pm0.04$ & $8.37\pm0.04$ & \\
This work (field) & $5-7$ & $7.0-9.0$ & $0.21\pm0.06$ & $7.87\pm0.06$ & \\
\hline
This work (all) & $5-7$ & $7.0-9.0$ & $0.25\pm0.01$ & $7.91\pm0.01$ & \multirow{3}{*}{R3 \citepalias{Sanders_24}} \\
This work (cluster) & $5-7$ & $7.0-9.0$ & $0.34\pm0.01$ & $8.14\pm0.01$ & \\
This work (field) & $5-7$ & $7.0-9.0$ & $0.20\pm0.05$ & $7.77\pm0.05$ & \\
\hline
\citet{Wang_22} (cluster) & $2.2$ & $9.0-10.4$ & $0.14\pm0.02$ & $8.44\pm0.18$ & R3, R2 \citepalias{Bian_18} \\
\hline
\citet{Chakraborty_24} & $3-10$ & $7.0-10.0$ & $0.21 \pm 0.03$ & $7.99 \pm 0.21$ & $T_e$ \\
\hline
\citet{Sarkar_25} & $4-10$ & $7.5-10.0$ & $0.27 \pm 0.02$ & $8.28 \pm 0.08$ & R3, O32 \citepalias{Curti_20} \\
\hline
\citet{Nakajima_23} & $4-10$ & $7.0-10.0$ & $0.25 \pm 0.03$ & $8.24 \pm 0.05$ & $T_e$, R23, O32 \citepalias{Nakajima_22} \\
\hline
\citet{Curti_24} & $3-10$ & $7.0-10.0$ & $0.17 \pm 0.03$ & $8.06 \pm 0.18$ & R3, R2 \citepalias{Curti_20} \\
\hline
\citet{Heintz_23} & $7-10$ & $7.5-10.0$ & $0.33$ & $7.95$ & R3, R2, R23 \citepalias{Sanders_24} \\
\hline
\citet{Chemerynska_24} & $6-8$ & $6.0-8.0$ & $0.39 \pm 0.02$ & $8.42 \pm 0.17$ & R3, R2, O32 \citepalias{Sanders_24} \\
\hline
\citet{Langeroodi_23} & $7-10$ & $7.0-10.0$ & $0.24 \pm 0.14$ & $7.92 \pm 0.18$ & R23, O32 \citepalias{Izotov_19} \\
\hline
\multirow{3}{*}{\citet{Sanders_21}} & $0$ & $8.7-11.5$ & $0.28 \pm 0.01$ & $8.77 \pm 0.01$ & $T_e$ \\
 & $2.3$ & $9.0-11.0$ & $0.30 \pm 0.02$ & $8.51 \pm 0.02$ & R3, O32, Ne3O2 \citepalias{Bian_18} \\
 & $3.3$ & $9.0-11.0$ & $0.29 \pm 0.02$ & $8.41 \pm 0.03$ & R3, O32, Ne3O2 \citepalias{Bian_18} \\
\hline
\multirow{2}{*}{\citet{Lim_23}} & $2$ & $6.5-9.5$ & $0.16 \pm 0.02$ & $8.50 \pm 0.13$ & \multirow{2}{*}{O32 \citepalias{Bian_18}} \\
 & $3$ & $6.5-9.5$ & $0.16 \pm 0.01$ & $8.40 \pm 0.06$ & \\
\hline
\multirow{2}{*}{\citet{He_24}} & $1.8-2.3$ & $6.9-10.0$ & $0.23 \pm 0.03$ & $8.54 \pm 0.12$ & \multirow{2}{*}{R3, R2 \citepalias{Bian_18}} \\
 & $2.6-3.4$ & $6.9-10.0$ & $0.26 \pm 0.04$ & $8.57 \pm 0.15$ & \\
\hline
\end{tabular}
\end{table*}

In Table \ref{tab:r3_info}, we provide the measured R3 ratios and the derived metallicities in different bins. In Table \ref{tab:mzr_info} we list our measured MZR properties in comparison with measurements in the literature.

\end{appendix}


\end{document}